\documentclass[11pt,a4paper]{article}

\usepackage[T1]{fontenc}
\usepackage[utf8]{inputenc}
\usepackage[english]{babel}
\usepackage{lmodern}
\usepackage[a4paper,margin=25mm]{geometry}
\usepackage{authblk}
\usepackage{booktabs}
\usepackage{enumitem}
\usepackage{graphicx}
\usepackage{xcolor}
\usepackage{hyperref}
\usepackage{xspace}
\usepackage{tabularx}
\usepackage{array}
\usepackage{ragged2e}
\usepackage{listings}
\usepackage[leftmargin=0.5cm,rightmargin=0.5cm,vskip=6pt]{quoting}
\usepackage[autostyle=true]{csquotes}
\usepackage[
  backend=biber,
  style=ieee,
  sorting=none,
  doi=true,
  url=true,
  eprint=false
]{biblatex}
\AtEveryBibitem{%
  \clearlist{language}%
  \clearfield{eprint}%
  \clearfield{eprinttype}%
  \clearfield{eprintclass}%
  \iffieldundef{doi}
    {}
    {\clearfield{url}\clearfield{urldate}}%
}

\SetBlockEnvironment{quoting}

\hypersetup{colorlinks=true,linkcolor=blue,citecolor=blue,urlcolor=blue}
\setlist{nosep}
\newcolumntype{C}{>{\centering\arraybackslash}X}
\newcolumntype{R}{>{\raggedleft\arraybackslash}X}
\newcolumntype{Y}{>{\RaggedRight\arraybackslash}X}
\newcommand{\cpntwo}{%
  \texorpdfstring{\textsc{CP}\textsuperscript{2}\textsc{N}\textsuperscript{2}}%
  {CP2N2}\xspace}
\def\BibTeX{{\rm B\kern-.05em{\sc i\kern-.025em b}\kern-.08em
    T\kern-.1667em\lower.7ex\hbox{E}\kern-.125emX}}

\title{\cpntwo: A Control Plane for Heterogeneous Physical Neural Networks}
\author[1]{Stefan Fischer\thanks{Corresponding author: stefan.fischer@uni-luebeck.de}}
\author[1]{Maliheh Hariri}
\author[2]{Sebastian Otte}
\affil[1]{University of L\"ubeck, Institute of Telematics, Ratzeburger Allee 160, 23562 L\"ubeck, Germany}
\affil[2]{University of L\"ubeck, Institute of Robotics and Cognitive Systems, Ratzeburger Allee 160, 23562 L\"ubeck, Germany}
\date{}

\begin{document}
\setlength{\emergencystretch}{3em}
\maketitle

\begin{center}
\small\itshape
This manuscript has been submitted for publication. It may be revised during peer review, and the copyright and licensing terms of any accepted or published version may differ from those of this preprint.
\end{center}

\begin{abstract}
Physical neural networks (PNNs) exploit diverse physical substrates---spanning molecular, wetware, and solid-state devices---to perform embodied neural computation close to physical processes. However, their operational heterogeneity in signal modalities, timing regimes, reset requirements, calibration lifecycles, and observability prevents their integration as interchangeable execution endpoints in edge, fog, and cloud workflows. We present \cpntwo, a substrate-aware control-plane architecture that bridges software-level orchestration and physical neural computing. \cpntwo introduces a three-plane model that decouples high-level orchestration, twin-state synchronisation, and substrate-specific execution under versioned Physical Neural Resource Contracts. It extends Model Context Protocol (MCP)-style discovery and invocation with explicit lifecycle management, atomic leases, telemetry freshness verification, and fail-closed safety boundaries. We evaluate a reference prototype across contrasting behavioural twins, a same-host multi-process deployment under fault injection, and the Cortical Labs SDK Simulator. In a 945-request concurrency campaign, \cpntwo enforces typed contention and fail-closed recovery with sub-millisecond local orchestration overhead. Furthermore, in a 160-decision Agent-to-PNN campaign with a hosted LLM planner, independent enforcement ensured that 100\% of executed actions remained strictly safe, verified, and reconciled despite planner-level inconsistencies. \cpntwo thus provides an enforceable systems abstraction for integrating scarce, stateful physical AI substrates into agentic computing environments.
\end{abstract}


\section{Introduction}
\label{sec:introduction}

Physical neural networks, referred to here as PNNs, perform neural computation
directly in physical substrates rather than only through numerical simulation
on conventional digital hardware. Their material diversity creates new
opportunities for placing computation close to sensing, actuation, and the
physical process itself~\cite{fischer2026beyondsilicon}. It also creates a
systems integration problem. PNNs differ in their I/O modalities, timing,
calibration, reset behaviour, and observability, so they cannot be managed as
interchangeable tensor-execution endpoints.

This becomes a systems issue when extreme-edge resources are integrated into
larger edge, fog, and cloud workflows rather than treated as isolated laboratory
demonstrations. A user should be able to state an application-level goal without
granting an agent authority over physical control primitives. A PNN-aware
application must instead determine which resources are suitable, whether their
current state is valid, and whether preparation, calibration, reset, or recovery
is required. The missing abstraction is therefore a control layer that exposes
the operational semantics of physical substrates to software-level resource
management.

Existing protocols and substrate-specific interfaces provide parts of this
functionality, but none defines how an agent-facing workflow should select and
supervise lifecycle-sensitive PNN resources under a shared operational
contract. \cpntwo addresses this gap while retaining MCP-style discovery and
invocation. It adds capability, timing, lifecycle, telemetry, and digital-twin
semantics and separates software-level orchestration, twin-state management,
and substrate-specific execution. The substrates may reside at the extreme
edge, while the control plane runs at the edge, fog, or cloud. This paper
evaluates a first reference architecture and its control abstractions, not the
performance of mature physical substrates or wide-area deployments.

The contributions are:
\begin{itemize}
    \item We frame heterogeneous PNN integration as a control-plane problem in which software must account for substrate-specific operational constraints.
    \item We define the three-plane \cpntwo architecture and a versioned capability model for signal modality, timing, lifecycle, observability, programmability, and training mode.
    \item We implement the architecture with twin-backed bindings, a distributed same-host path, the Cortical Labs simulator, and the BioPattern Gate application workflow.
    \item We evaluate contract portability, admission, selection, recovery, audit, and overhead and conduct a confirmatory Agent-to-PNN campaign that separates probabilistic planning from independent enforcement.
\end{itemize}

Sections~\ref{sec:background}--\ref{sec:bindings} motivate, specify, and implement the architecture. Section~\ref{sec:evaluation} evaluates it, after which Sections~\ref{sec:related-work} and~\ref{sec:conclusion} position and conclude the work.

\section{Motivating Scenarios and System Context}
\label{sec:background}

\subsection{Autonomous Functional-Neurobiology Campaign}

Consider a scientific agent coordinating a multi-session campaign over scarce
living neural cultures. The user asks it to compare approved conditions, use
only validated assay presets, balance sessions across available cultures, stop
on safety or data-quality failures, and schedule a confirmatory follow-up only
if a pre-registered threshold is met.

The agent operates at campaign level. It may select an admissible resource, published assay, session order, and permitted follow-up, but it may not design
stimulation primitives, change endpoints after observing results, or bypass
human and laboratory approval. The systems pressure comes from scarcity,
exclusive access, calibration, age and experimental history, telemetry
freshness, recovery after ambiguous failures, and cross-session provenance.

Existing platforms for closed-loop and remote access to living neural systems make this a technically grounded near-future scenario \cite{kagan2022dishbrain,jordan2024neuroplatform}, even though this campaign has not yet been executed in reality.

\subsection{Adaptive Embodied System over Heterogeneous PNN Resources}

Consider an embodied or edge application that must maintain an event-stream
task under latency, locality, energy, and evidence constraints. A supervisory
agent may request an available physical-neural resource, validate its current
calibration, and select a safe fallback when health, drift, or telemetry
freshness violates policy. Candidate resources may span neuromorphic devices,
memristive or photonic reservoirs, biological systems, and simulators used
only at their declared evidence levels \cite{fischer2026beyondsilicon,pedersen2024nir}.

The time-critical encoder, inference/readout, and actuator loop remain in the
substrate-local runtime. Remote agent actions are deliberative: discovery,
admission, reservation, preparation, recalibration requests, and fallback.
This separation motivates a control plane above substrate-specific runtimes
without suggesting that an LLM or MCP client performs hard real-time control.

\subsection{PNNs as Heterogeneous Physical AI Substrates}

The two scenarios illustrate why PNNs cannot be treated as a single hardware
class. They comprise computational sub\-stra\-tes based on very different physical
mechanisms. Recent survey work places DNA strand-displacement systems,
reaction--diffusion media, living neural tissue, memristive and ferroelectric
in-memory devices, photonic circuits, and mechanical, microfluidic, or
iontronic architectures under this umbrella~\cite{fischer2026beyondsilicon}.
Across these substrates, familiar neural functions such as weighted propagation,
nonlinearity, memory, and adaptation are realised through different carriers,
state variables, and engineering constraints.

This diversity is central to the appeal of PNNs for embedded AI. Depending on the substrate, computation can be placed close to sensing, coupled tightly to the underlying physics of the application domain, or carried out under latency, energy, bandwidth, or environmental constraints for which conventional accelerators are a poor fit. Many of the most compelling deployment scenarios, therefore, place PNNs at the extreme edge, i.e., at the physically immediate boundary to the sensed or influenced process. The potential of PNNs lies largely in substrate diversity rather than in uniformity.

At the same time, this diversity makes integration difficult. A PNN is a physical resource with its own I/O modality, timing behaviour, calibration requirements, reset conditions, and observability limits. A photonic substrate, a DNA-based circuit, and a wetware system may all implement neural computation, but they differ substantially in how they are configured, when their outputs can be trusted, how drift appears, and what a supervising software stack can realistically observe or control. Once such substrates are treated as deployable resources in edge, fog, and cloud hierarchies rather than as isolated laboratory demonstrations, these differences become a systems concern.

\subsection{MCP as a Baseline Discovery and Invocation Model}

One useful reference point for this systems concern is the Model Context
Protocol, or MCP. At a high level, MCP defines a standardised way to describe
external capabilities in machine-readable form and to make them discoverable
and invocable by AI systems~\cite{mcp2025spec}. Its contribution is therefore
not a model representation, compiler, or runtime. Instead, it provides a common
interaction structure for exposing resources and tools to software.

This abstraction matches how many current AI systems are assembled: they are no longer built as monolithic applications, but as orchestrated workflows in which an agent or supervisory component inspects available capabilities at runtime, selects a suitable external resource, and integrates the returned result into a larger reasoning loop. In that setting, explicit and machine-readable resource descriptions are preferable to hidden, application-specific integrations.

For PNN integration, MCP is relevant because it already captures two properties that are also needed here. First, heterogeneous resources should be discoverable rather than hard-coded. Second, invocation should proceed through a uniform software-visible interface rather than through ad hoc glue logic. For physical AI systems, these two properties cover the software-facing part of integration but not the substrate-facing one. However, MCP does not by itself express the physical semantics that determine whether a substrate is actually usable for a given task.

\subsection{Why Existing Interfaces Still Leave an Integration Gap}

Existing interfaces cover important parts of this problem without providing a
shared operational model across substrate families. NIR and EdgeMap address
representation and backend mapping, while BioCRNpyler and wetware platforms
provide family-specific compilation or experiment interfaces
\cite{pedersen2024nir,xue2023edgemap,poole2022biocrnpyler,kagan2022dishbrain,jordan2024neuroplatform}.
They do not jointly expose reset, calibration, lifecycle, telemetry, and twin
validity across chemical, biological, and device-based resources; detailed
comparisons are deferred to Section~\ref{sec:related-work}.

From the viewpoint of an embedded or distributed application, a PNN-facing interface must expose signals, latency, initialisation and recalibration, reset or replacement, runtime programmability, and telemetry for detecting drift or invalid state. Existing interfaces usually expose only part of this picture within one substrate family.

Conventional software stacks, therefore, still lack a proper control-plane abstraction for heterogeneous PNNs. What is needed is an orchestration layer that keeps substrate differences explicit and makes them usable for software-level discovery, matching, invocation, supervision, and digital-twin coupling. This is the design space in which we position \cpntwo.

\section{Requirements, System Model, and Threat Model}
\label{sec:requirements}

The gap identified above is a systems problem: heterogeneous PNN substrates can only be used reliably if their relevant physical and operational properties are exposed at the control-plane level. From the viewpoint of an orchestrator or supervisory application, the control plane must therefore support discovery, selection, invocation, monitoring, and lifecycle management in a substrate-aware manner.

\subsection{Requirements}

\emph{R1: Capability discovery and task-to-substrate matching.}
A PNN resource must be discoverable by more than endpoint presence. Software needs machine-readable information about which computational functions a substrate supports, under which input/output modalities, and under which operational conditions. This information must also support principled task-to-substrate matching: an orchestrator should be able to decide which substrate fits which task under constraints such as modality, latency, energy, autonomy, or lifecycle cost. Without such structured descriptors, placement remains ad hoc.

\emph{R2: Typed multi-physics I/O descriptions.}
Many PNNs do not simply consume and produce tensors. Their interfaces may involve molecular concentrations, spike trains, optical intensities, conductance states, or mechanical excitation patterns. The control plane must therefore describe I/O in typed, substrate-aware terms, including carrier, encoding, admissible ranges, sampling assumptions, and required transduction steps.

\emph{R3: Time and synchronisation semantics.}
Timing is intrinsic to PNN behaviour. Substrates differ widely in latency, observation windows, settling behaviour, and freshness constraints. A usable control plane must expose these semantics explicitly so that orchestration and digital-twin coupling can judge whether results are still valid for the task at hand.

\emph{R4: Lifecycle and reset control.}
Many substrates are not always-ready resources. They may require warm-up, priming, flushing, calibration, thermal cycling, or replacement, while others support rapid reset or near-continuous reconfiguration. The control plane must therefore represent lifecycle states explicitly, rather than collapsing them into a simple notion of availability.

\emph{R5: Telemetry and digital-twin coupling.}
A physical substrate requires more observability than a final output alone. The control plane should expose telemetry for supervision, debugging, recalibration, and failure analysis, while also providing clear bindings to digital twins. Relevant metadata includes calibration timestamps, divergence indicators, and estimates of twin validity.

\emph{R6: Training- and programmability-aware operation.}
PNNs differ substantially in configurability. Some are fixed after ex-situ training, others allow limited retuning, hybrid update procedures, or in-materio adaptation. The control plane must make these distinctions explicit, since they affect deployment, rollback, validation, reproducibility, and update handling.

\emph{R7: Safety, isolation, and tenancy constraints.}
Physical substrates may be fragile, hazardous, biologically sensitive, expensive, or shared. The control plane must therefore support policies for concurrent access, admissible stimulation ranges, cooldown periods, hygiene or biosafety constraints, and authorisation. A shared PNN cannot be exposed as an unconstrained stateless service.

Table~\ref{tab:scenario-traceability} links the two motivating scenarios to
the mechanisms evaluated in the systems core.

\begin{table*}[t]
    \scriptsize
    \centering
    \caption{Scenario-to-requirement and mechanism traceability.}
    \label{tab:scenario-traceability}
    \begin{tabularx}{\textwidth}{@{}p{0.07\textwidth}p{0.22\textwidth}p{0.34\textwidth}X@{}}
        \toprule
        \textbf{Req.} & \textbf{Scenario pressure} & \textbf{\cpntwo mechanism} & \textbf{Evaluation locus} \\
        \midrule
        R4, R7 & scarce cultures or exclusive devices & leases, ownership, and expiry & lifecycle tests and concurrent-client campaign \\
        R3--R5 & changing health, calibration, and drift & telemetry provenance, freshness, and fail-closed admission & contract, policy, and stale-state tests \\
        R7 & unsafe agent-generated physical parameters & server-owned assays, least authority, and external approval & MCP-schema and bypass tests \\
        R1, R2, R6 & heterogeneous modalities and substrates & resource contract and adapter capabilities & conformance and portability evaluation \\
        R3, R4, R7 & remote timeout with uncertain state & explicit abort, status unknown, and reconciliation & fault and recovery campaign \\
        R3 & local real-time loop versus remote deliberation & control-adapter/substrate-runtime split & adapter conformance and deployment architecture \\
        R5, R7 & cross-session accountability & audit chain, manifests, artefacts, and checksums & trace and artefact audit \\
        \bottomrule
    \end{tabularx}
\end{table*}

\subsection{System Model}

The system contains an untrusted or fallible agent client, a policy-enforcing
\cpntwo control plane, a registry of versioned resource contracts, control
adapters, and substrate-local runtimes. A resource may be exclusive, remote,
temporarily unreachable, or only partially observable. Dynamic facts carry
source and observation time; unknown state remains unknown. The control plane
performs deliberative orchestration, while time-critical physical I/O remains
inside the local runtime. Evidence level is an attested or configured runtime
property and cannot be selected by the agent.

\subsection{Threat Model and Trust Boundaries}

We assume that agent output may be malformed, adversarial, stale, or
overconfident; clients may race for an exclusive resource; messages may be
lost, delayed, duplicated, or reordered; and a timeout may leave provider
state uncertain. The control plane, policy store, assay catalog, lease store,
approval verifier, and audit chain form the trusted computing base. Adapters
and provider telemetry are authenticated inputs but may fail or become stale.
The design therefore validates every state-changing request, separates hard
admission constraints from ranking, binds execution to a lease and
idempotency key, exposes explicit abort and reconciliation, and never promotes
a simulator label to physical hardware. Compromise of the host, provider, or
substrate firmware and the physical correctness of provider telemetry remain
outside the present threat model.

Overall, a PNN control plane must provide more than transport-level interoperability. Its role is not to hide substrate differences, but to expose them in a structured form that software can use for discovery, selection, invocation, supervision, and lifecycle management. The next section presents the \cpntwo architecture built around this idea.

\section{\cpntwo Control-Plane Design and Architecture}
\label{sec:architecture}

\begin{figure}[t]
    \centering
    \includegraphics[width=\linewidth]{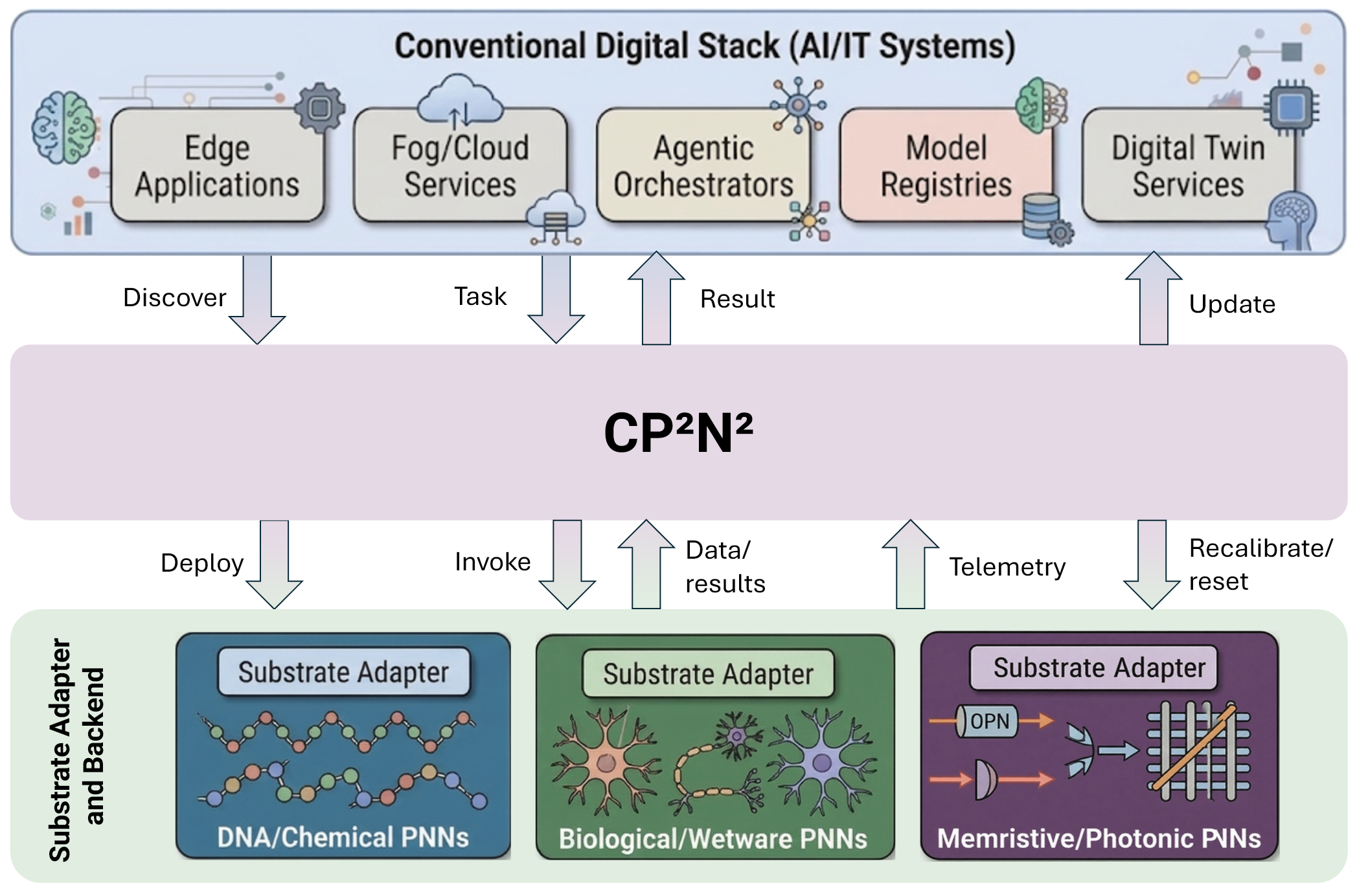}
    \caption{High-level system architecture of \cpntwo, connecting conventional AI/IT systems to heterogeneous physical-neural substrates through a control plane and substrate adapters. The component artwork was generated with Gemini NanoBanana (web service, accessed 27 April 2026; exact model revision not exposed); a label was corrected with OpenAI's built-in image-editing service (accessed 30 July 2026; exact model revision not exposed).}
    \label{fig:cp2n2-highlevel}
\end{figure}

Figure~\ref{fig:cp2n2-highlevel} places \cpntwo within the overall system stack and shows its role as the substrate-aware interface between conventional digital AI/IT components and heterogeneous physical neural substrates, many of which may naturally reside at the extreme edge. At this level, \cpntwo appears as the software layer through which edge, fog, or cloud systems discover available substrate capabilities, submit execution requests, receive runtime information, and exchange state with digital-twin services. This view clarifies the architectural role of \cpntwo: it is not a substrate-specific runtime, but an orchestration layer between software-level intent and materially embodied computation.

\begin{figure*}[t]
    \centering
    \includegraphics[width=\linewidth]{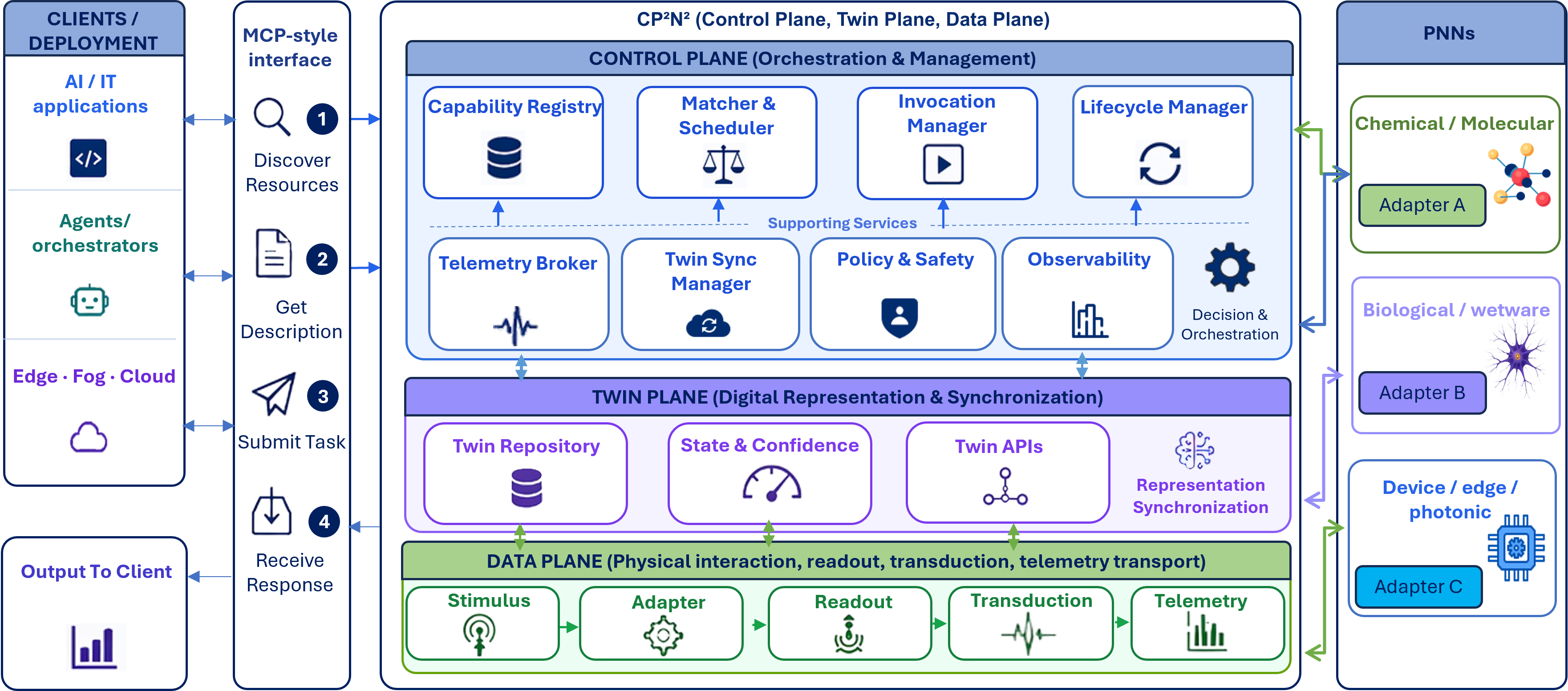}
    \caption{Internal three-plane architecture of \cpntwo. \cpntwo separates control-plane orchestration, twin-plane state and validity management, and data-plane substrate access. The control plane handles capability discovery, task-to-substrate matching, invocation, lifecycle control, and policy enforcement. The twin plane aggregates telemetry from substrate adapters, maintains synchronised substrate state, and estimates calibration validity and drift. The data plane contains substrate-specific adapters that execute physical I/O and readout for heterogeneous PNN backends.
    }
    \label{fig:cp2n2-three-plane}
\end{figure*}

We now move from this external placement view to the internal organisation of \cpntwo. We do not treat \cpntwo as a single middleware block, but instead decompose it into three coordinated planes that separate orchestration, twin-state management, and substrate access. Figure~\ref{fig:cp2n2-three-plane} summarises this internal structure and the main control, telemetry, and feedback paths across it.

\subsection{Control Plane, Twin Plane, and Data Plane}

The main architectural decision in \cpntwo is to separate software-facing control logic from substrate-facing execution and from the management of twin-visible runtime state. This yields a three-plane design. The Control Plane handles soft\-ware-le\-vel decisions and session coordination. The Twin Plane maintains synchronised, validity-aware state associated with the substrate. The Data Plane performs the actual interaction with physical or emulated backends. This split keeps responsibilities clear without isolating the planes from one another: invocation decisions depend on twin state, twin state depends on telemetry, and lifecycle actions ultimately affect substrate access.

\textbf{Control plane.} The control plane is the entry point for applications, orchestrators, and supervisory agents. It handles discovery, capability negotiation, policy enforcement, invocation, and lifecycle management. When a request is capability-driven, the control plane selects a suitable substrate from the available candidates. When a client already addresses a specific backend, the same plane validates feasibility, policy, interface, and readiness constraints for that target. In both cases, the control plane does not execute physical computation itself; it prepares, authorises, and supervises the corresponding session.

\textbf{Twin plane.} The twin plane maintains the digital representation associated with a substrate. Depending on the backend, this may take the form of a high-fidelity simulator, a reduced behavioural model, a calibration record, or only a best-effort validity estimate. The twin plane tracks synchronisation metadata, confidence, and drift-related status. This separation matters because the twin is not the substrate itself. Its value depends on how current it is, how well it matches observed behaviour, and whether the surrounding software can still rely on it for scheduling or control decisions.

\textbf{Data plane.} The data plane comprises the substrate-specific execution path. It includes stimulation, actuation, sensing, readout, and low-level telemetry transport between adapters and the physical or emulated backend. This plane is deliberately not uniform across substrates. A chemical backend may use concentration-valued inputs and fluorescence readouts, whereas a wetware backend may exchange stimulation patterns and spike recordings. \cpntwo does not hide these differences. It makes them accessible through software-visible descriptors and contracts so that higher-level systems can reason about them.

The result is a control structure in which orchestration remains software-visible, twin maintenance remains explicit, and physical execution remains substrate-specific. This is the layer that is missing in conventional software stacks when materially diverse PNNs have to be integrated into a larger system.

\subsection{Core Components}

Figure~\ref{fig:cp2n2-three-plane} sketches the main building blocks. In the architecture proposed here, the control and twin planes are realised through the following core services.

\textbf{Capability registry.} The registry stores descriptors for known PNN resources and their exposed capabilities. These descriptors support discovery queries such as ``find a substrate that accepts spike-like event input and supports low-latency repeated invocation'' or ``find a substrate that supports in-sample molecular processing under slow assay semantics.''

\textbf{Task-to-substrate matcher.} Based on resource descriptors and current operating state, this component ranks candidate substrates when a request does not yet specify a concrete backend. It is the point at which capability compatibility, timing constraints, deployment locality, lifecycle cost, and twin state become operational selection criteria. If a client already targets a known substrate, the role of this component reduces to compatibility, readiness, and admission checks rather than open-ended selection.

\textbf{Invocation manager.} The invocation manager turns a request into a concrete session. It establishes the relevant execution context, negotiates timing, lifecycle, and telemetry expectations, activates the appropriate adapter, and tracks whether a request is running, paused, completed, rejected, or invalidated.

\textbf{Lifecycle manager.} This component supervises warm-up, priming, calibration, reset, cooldown, recovery, and related transitions. For physical substrates, these state changes are often as important as the compute step itself, especially in slow, fragile, or drift-prone systems.

\textbf{Telemetry broker.} \cpntwo collects runtime signals that matter for control and supervision. Depending on the substrate, this may include final outputs, intermediate observations, health indicators, calibration state, or drift warnings. These signals are forwarded to local or remote consumers as needed and also provide input to the twin plane.

\textbf{Policy and safety manager.} The policy layer enforces admissible operating regions, authorisation, tenant isolation, and substrate-specific safety rules. This becomes essential as soon as PNN resources are shared, fragile, hazardous, or biologically sensitive.

\textbf{Twin synchronisation manager.} This manager associates telemetry with the corresponding twin state and updates synchronisation metadata. It can flag stale twin state, unexpected behavioural deviation, or situations in which additional measurements are required before the resource should be used again.

\subsection{Admission and Selection}

Selection in \cpntwo is a three-stage decision. \emph{Admission} evaluates hard
task, modality, evidence, supervision, safety, and policy constraints.
\emph{Feasibility} then evaluates current availability, reservability,
telemetry freshness, and budget. Only candidates that pass both stages enter
\emph{ranking}. A directed request still passes admission and feasibility; it
merely removes the final choice among multiple candidates.

The principal ranking is lexicographic and uses an explicit policy profile,
for example safety/\allowbreak{}freshness first, latency first, or locality/\allowbreak{}cost first.
This prevents a favourable soft score from compensating for a failed hard
constraint. Each decision records the selected policy, ordered comparison
keys, observed values, and a reason for every exclusion. A transparent
weighted heuristic is retained only as an evaluation baseline, together with
static-priority, constraint-only, and random-admissible selectors. With \cpntwo, we do not aim for universal optimality; we want to achieve reproducible decisions whose hard constraints and preferences can be inspected independently.

\subsection{End-to-End Workflow}

A typical \cpntwo workflow starts in one of two ways, comparable to using yellow and white pages, respectively, for the detection of providers. In a capability-driven
workflow, a client first discovers suitable resources and then submits a request
at the level of desired functionality and constraints. In a directed workflow,
the client already names a specific substrate and asks the control plane to
validate and invoke it. Consider a biological PNN that accepts a published
stimulation assay and returns time-windowed observations together with health
and provenance metadata.

In the capability-driven variant, the client discovers resources that advertise
the required stimulation, observation, timing, and telemetry capabilities. In
the directed variant, it addresses one known biological resource. In both cases,
the request names only a published assay and admissible constraints. Low-level
stimulation parameters, safety policy, evidence labels, and approval remain
server controlled. The matcher evaluates static descriptors together with
dynamic state such as readiness, health, drift, and telemetry freshness. The
control plane then prepares the selected adapter and establishes the execution,
lifecycle, lease, and audit context.

The return path is equally important. The selected adapter interacts with the
backend and receives backend-native observations together with runtime metadata.
The control plane validates the declared postconditions, associates the result
with the current lease and lifecycle state, and returns a normalised response
that exposes only the fields permitted by policy. A result can therefore carry
observations, provenance, validity, and recovery information without exposing
uncontrolled physical output or silently converting an uncertain state into a
successful one.

The interaction style remains close to standard MCP usage. A client discovers
resources or addresses a known one, submits a structured request through a common
interface, receives a normalised response, and can inspect machine-readable
auxiliary state. What changes in \cpntwo is the semantics of that state: timing,
resetability, observability, evidence status, and recovery actions become
explicit software-visible concerns rather than remaining hidden inside
substrate-specific glue code.

We will show a concrete version of this workflow along with a visual description in Section~\ref{sec:sample-workflow}.

\subsection{Physical Neural Resource Contract}
\label{sec:capability-model}

The Physical Neural Resource Contract
is not merely a catalogue entry or a passive description of a backend. It is a
versioned, machine-checkable agreement between the client-facing control plane, the substrate adapter, and the resource runtime. The contract declares
preconditions for admission and execution, invariants that must hold during a
leased session, postconditions for accepting a result, and explicit outcomes
when an obligation cannot be met. It also specifies available capabilities,
valid state transitions, required telemetry, and operating limits. These fields
are checked by the control plane during admission, execution, and validation.
The contract therefore acts as both an interface specification and an
enforceable operational boundary.

The contract distinguishes between two related kinds of description. A
\emph{resource descriptor} identifies a concrete substrate instance and its
operating context. A \emph{capability descriptor} states what that resource can
do and under which conditions it can be used. Keeping both in one versioned
contract lets the control plane validate a request against stable capabilities
and current admissibility without treating either as sufficient on its own.

The resource descriptor captures relatively stable information such as substrate class, physical location, adapter type, tenancy constraints, and the identifier of the associated twin model. It describes what kind of resource is being exposed and under which operational conditions it exists.

The capability descriptor is closer to execution. It records the properties that matter when software has to decide whether and how a resource can be used: signal modality, admissible I/O types, timing regime, programmability, observability, lifecycle affordances, and telemetry availability. These are the computational and operational semantics that the control plane must see.

Table~\ref{tab:capabilities} summarises the descriptor categories used in our initial model.

\begin{table}[t]
    \centering
    \caption{Core categories in the \cpntwo capability model.}
    \label{tab:capabilities}
    \begin{tabular}{@{}p{0.28\columnwidth}p{0.64\columnwidth}@{}}
        \toprule
        \textbf{Category} & \textbf{Representative fields} \\
        \midrule
        Substrate identity & substrate class, adapter type, location, twin binding \\
        Signal semantics & input/output modality, encoding, admissible ranges, transduction needs \\
        Timing semantics & latency regime, observation window, freshness expectations, trigger mode \\
        Lifecycle semantics & warm-up, resetability, calibration need, recovery and cooldown states \\
        Programmability & fixed, configurable, tunable, in-situ adaptive \\
        Observability & output channels, internal telemetry, drift indicators, twin confidence \\
        Policy and tenancy & exclusivity, safety bounds, authorisation, concurrency limits \\
        \bottomrule
    \end{tabular}
\end{table}

These fields are not passive documentation. They are ma\-chi\-ne-readable inputs to
matching, admission control, invocation setup, and supervision. A scheduler may
reject a candidate whose contract signals assay-style timing when the request
expects repeated low-latency invocations. A policy layer may refuse access when
exclusivity, safety, or biosafety constraints are incompatible with the request.
Because the contract is versioned, changes to capabilities, lifecycle semantics,
telemetry obligations, or error meanings can be detected explicitly instead of
being inferred from adapter-specific behaviour. The contract therefore provides
the stable control surface needed for portability across heterogeneous PNNs.

\subsection{Lifecycle Protocol, Leases, Recovery, and Error Semantics}

\begin{figure*}[t]
    \centering
    \includegraphics[width=\textwidth]{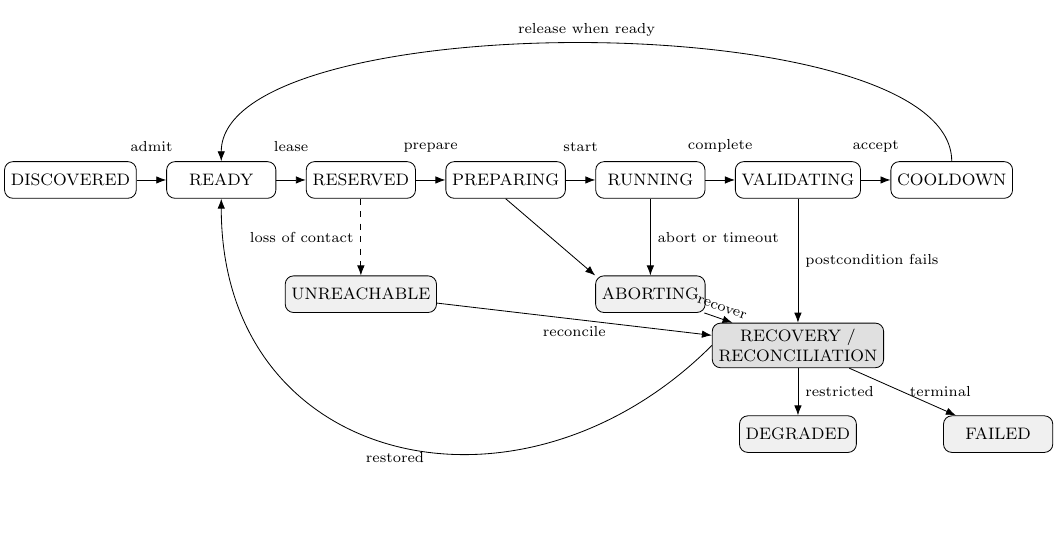}
    \caption{Abstract \cpntwo lifecycle state machine. The normal path prepares,
    executes, validates, cools down, and releases a leased resource. Aborts,
    failed postconditions, and loss of contact lead to explicit exceptional
    states. Recovery or reconciliation establishes whether the resource can
    return to \texttt{READY} or must remain degraded or failed.}
    \label{fig:lifecycle-state}
\end{figure*}

Descriptors alone are insufficient because admissibility changes during a
session. Figure~\ref{fig:lifecycle-state} summarises the lifecycle protocol and
its fail-closed recovery paths. \cpntwo uses the common lifecycle
\texttt{DISCOVERED}$\rightarrow$\allowbreak{}\texttt{READY}$\rightarrow$\allowbreak{}\texttt{RESERVED}
$\rightarrow$\allowbreak{}\texttt{PREPARING}$\rightarrow$\allowbreak{}\texttt{RUNNING}
$\rightarrow$\allowbreak{}\texttt{VALIDATING}$\rightarrow$\allowbreak{}\texttt{COOLDOWN}
$\rightarrow$\allowbreak{}\texttt{READY}. Active states can enter \texttt{ABORTING} and
then \texttt{DEGRADED}, \texttt{FAILED}, or \texttt{READY}; loss of contact
produces \texttt{UNREACHABLE}, never implicit success. State-changing
operations carry an expected state version, correlation ID, and idempotency
key.

Exclusive resources require an atomic, owner-bound, time-limited lease before
preparation. Renewal and release verify ownership; expiry prevents later
execution even if a stale client continues. Separate preparation, execution,
validation, abort, and reconciliation timeouts preserve the phase in which
uncertainty arose. After a timeout, \cpntwo queries provider state and records
\texttt{EXECUTION\_STATUS\_UNKNOWN} until reconciliation establishes an
authoritative outcome.

The telemetry contract attaches value, unit, source, observation time, receipt
time, uncertainty, and validity period to dynamic facts. Admission uses the
observation and validity interval rather than a client-supplied ``fresh''
flag. Missing safety information remains unknown and fails closed. Errors such
as \texttt{INADMISSIBLE}, \texttt{POLICY\_DENIED},
\texttt{RESOURCE\_BUSY}, \texttt{LEASE\_EXPIRED},
\texttt{TELEMETRY\_STALE}, and \texttt{POSTCONDITION\_FAILED} are stable
control-plane outcomes rather than adapter-specific strings.

These semantics extend MCP-style discovery and invocation with the
domain-specific state required for embodied computation \cite{mcp2025spec}.
They make physical reset cost, exclusivity, observability, telemetry validity,
and uncertain post-failure state visible to the orchestrator without moving
time-critical control out of the substrate-local runtime.

\section{Implementation and Adapter Integration}
\label{sec:bindings}

We now move from the abstract capability model to its realisation in the
reference prototype. To avoid conflating an intended physical substrate with
the software used in this paper, we distinguish three terms. A \emph{target
substrate class} denotes the physical technology that \cpntwo is designed to
support. A \emph{binding} maps the common resource contract to the semantics of
that class. A \emph{runtime} is the concrete mock, twin, service, simulator, or
physical resource that is actually executed. The prototype bindings were
chosen to exercise contrasting signal modalities, timing regimes, lifecycle
costs, telemetry, and deployment boundaries. None of the three core bindings
executes its corresponding physical substrate in this work.

We classify every runtime using the provenance categories we describe in
Table~\ref{tab:evidence-levels}. These categories are not an external standard
and do not form a cumulative maturity score. They identify what kind of system
produced an observation and therefore which conclusions that observation can
support. In particular, repeated runs at E0--E4 cannot establish an E5 claim
about physical hardware or living neural tissue.

\begin{table}[t]
    \scriptsize
    \centering
    \caption{Runtime-provenance categories used in this paper.}
    \label{tab:evidence-levels}
    \begin{tabularx}{\columnwidth}{@{}p{0.08\columnwidth}p{0.27\columnwidth}X@{}}
        \toprule
        \textbf{Class} & \textbf{Runtime provenance} & \textbf{Role in this work} \\
        \midrule
        E0 & interface mock & unit, schema, and error-path tests \\
        E1 & behavioural twin & core binding, policy, orchestration, and local overhead tests \\
        E2 & same-host service & process boundary, controlled faults, and service-path overhead \\
        E3 & official SDK simulator & provider integration, recording, reconstruction, and replay without physical execution \\
        E4 & independent remote simulator & defined for remote-path evidence but not instantiated here \\
        E5 & physical hardware or living tissue & required for physical or biological claims but not instantiated here \\
        \bottomrule
    \end{tabularx}
\end{table}

\subsection{DNA/Chemical Binding}

The DNA/chemical binding maps concentration-driven, assay-style computation to
an E1 ODE-based behavioural twin and a chemical adapter; no molecular reaction
is executed. Its contract describes concentration-valued I/O, slow convergence,
\texttt{flush} and \texttt{recharge} reset modes, and telemetry for
contamination, calibration, convergence, and drift. These properties make the
binding a lifecycle test: execution may require preparation, delayed readout,
validity checks, and cleanup before the resource can be admitted again.

An autonomous water-monitoring station illustrates the intended operational
setting. DNA neural systems have demonstrated molecular pattern classification,
although not this deployment \cite{cherry2025dnann}. A local classifier could
map concentration-encoded samples to a risk category for confirmatory analysis.
\cpntwo would admit the assay only while reagent state, contamination telemetry,
calibration confidence, and twin validity satisfy the contract, and would track
convergence and the subsequent flush or recharge cycle.

\subsection{Biological/Wetware Binding}

The biological/wetware binding shifts the emphasis from slow equilibration to
repeated interaction with a state-sensitive resource. An E1 synthetic
spike-response twin and wetware adapter expose spike-oriented input,
millisecond-scale timing, viability-sensitive state, \texttt{rest} and
\texttt{recalibrate} operations, and telemetry for firing rate, response delay,
noise, viability, and drift. No living tissue is executed. The shared interface
is unchanged, but session validity now depends primarily on health and
closed-loop observability.

The prototype also includes an E3 integration with the official Cortical Labs
SDK Simulator~\cite{hogan2026clapi}. Its two surfaces cover control-plane
operation and the documented provider API and recording path; both remain
software-only\footnote{As of the date of writing this article, it is still
difficult to get access to real-world PNNs. The current wait time for access to
a Cortical Labs CL1 neural computer through the Cortical Labs Cloud is roughly
half a year.}. A motivating physical use case is an agent-operated
neuropharmacology laboratory. Multi-well microelectrode-array cultures are
already used to measure compound-induced changes in neuronal network function
\cite{strickland2018toxcast}. Before granting exclusive access to a published
pre/post-exposure assay, \cpntwo checks culture identity, viability,
supervision, calibration, and recovery intervals; out-of-contract health or
drift telemetry causes rejection or rescheduling.

\subsection{Memristive/Photonic Binding}

The memristive/photonic binding represents a more device-like operating regime.
Its E1 vector/\allowbreak{}tensor-oriented twin supports repeated millisecond-scale
invocation and reports drift, latency, and an energy proxy; it executes no
memristive or photonic device. Calibration age, reprogramming overhead, and
locality remain visible even though the interface resembles that of a
conventional accelerator. In the prototype, this binding drives the fallback
and drift-recovery tests.

Real-time signal equalisation in an optical transceiver provides a concrete
deployment context; the task has recently been demonstrated on an integrated
silicon-photonic reservoir computer \cite{vanassche2026equalization}. A
supervisory agent can request a suitable local preset when link monitoring
detects dispersion or nonlinear distortion. \cpntwo then verifies calibration,
reserves and configures the device, and monitors drift and temperature. Loss of
admissibility redirects the workflow to conventional digital signal processing.

\subsection{Comparison of the Prototype Bindings}

Table~\ref{tab:bindings-overview} separates each target class or execution path from the runtime that represents it in the prototype. The goal is contrast rather than broad substrate coverage. The three E1 bindings exercise slow chemical dynamics, state-sensitive wetware semantics, and a low-latency device-like profile. The E2 path externalises the latter profile behind a same-host service boundary, while the E3 path adds an official provider simulator. Together, these runtimes form a small but deliberately diverse systems test bed.

\begin{table*}[t]
    \scriptsize
    \centering
    \caption{Prototype bindings and their runtime-provenance classification. The evidence class describes the runtime executed in this work, not the corresponding physical target substrate.}
    \label{tab:bindings-overview}
    \setlength{\tabcolsep}{5pt}
    \begin{tabularx}{\textwidth}{@{}p{0.17\textwidth}p{0.25\textwidth}p{0.08\textwidth}X@{}}
        \toprule
        \textbf{Target class or path} & \textbf{Prototype runtime} & \textbf{Class} & \textbf{Representative semantics} \\
        \midrule
        DNA/chemical & ODE-based behavioural twin with chemical adapter & E1 & concentration I/O, slow convergence, flush, recharge, and recalibration \\ \addlinespace
        Biological/wetware & synthetic spike-response twin with wetware adapter & E1 & stimulation and spike I/O, viability, rest, and recalibration \\ \addlinespace
        Memristive/photonic & local vector/\allowbreak{}tensor-oriented twin & E1 & low-latency invocation, drift, reset, and reprogramming \\ \addlinespace
        Externalised fast path & same device-like profile behind a same-host HTTP service & E2 & process boundary, controlled transport faults, and service-path overhead \\ \addlinespace
        Cortical Labs integration & official CL SDK Simulator & E3 & provider API, recording, reconstruction, packaging, and replay without physical execution \\
        \bottomrule
    \end{tabularx}
\end{table*}

E0 mocks support unit and error-path tests but do not constitute another
binding. Neither an E4 remote simulator nor an E5 physical PNN is executed.
This is consistent with the present empirical goal of validating control-plane
abstractions across controlled runtime regimes; physical-hardware validation
remains necessary before making real-world PNN claims.

\subsection{Reference Prototype}
\label{sec:prototype}

We implemented \cpntwo as a Python-based reference prototype for integrating heterogeneous physical AI bindings into edge-oriented software systems. The evaluated prototype uses one logical control-plane instance; lifecycle, lease, idempotency, run, and approval state are maintained by in-process stores, while principals and scopes are injected through server configuration rather than a production identity provider. The implementation is intentionally small and mostly single-machine. Its purpose is to validate the control-plane abstractions and their applicability to extreme-edge resources in broader software-managed workflows, rather than to provide a distributed production runtime.

The task model combines static descriptors with runtime state. A request can
specify a runtime preference, a maximum twin age, required telemetry, and the
availability of human supervision; the matcher evaluates these fields against
snapshots such as \texttt{health\_status}, \texttt{drift\_score}, and
\texttt{age\_of\_information\_ms}. After invocation, the orchestrator checks
postconditions and may reroute following preparation, invocation, telemetry, or
validity failures. An HTTP-backed adapter introduces a same-host process
boundary for the service-path experiments. The resulting implementation covers
local twin-backed bindings, an externalised service, and a provider simulator.

\subsection{BioPattern Gate Application Workflow}
\label{sec:sample-workflow}

\begin{figure*}[!t]
    \centering
    \includegraphics[width=\textwidth]{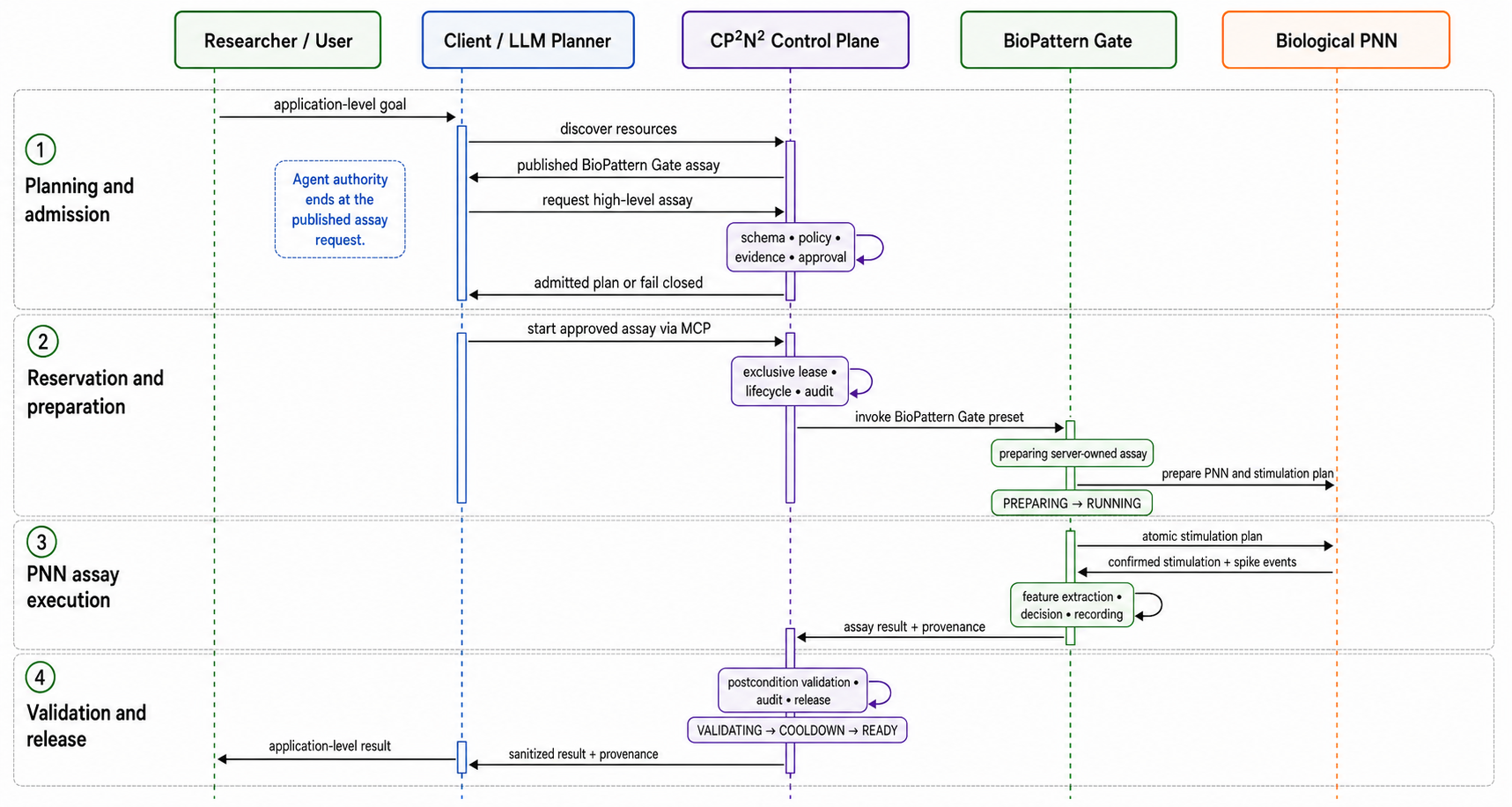}
    \caption{Conceptual BioPattern Gate sequence workflow for an integrated
    Agent-to-PNN scenario. The client uses MCP to request a published assay.
    \cpntwo independently checks admissibility, obtains an exclusive lease,
    advances the resource lifecycle, and invokes the BioPattern Gate preset.
    BioPattern Gate prepares the server-owned stimulation plan, conducts the
    assay on the PNN, extracts features, and returns the result and provenance.
    The control plane then validates postconditions, records the audit trail,
    releases the resource, and returns a sanitised result to the client.
    The figure was generated with OpenAI's built-in image-generation service,
    accessed 3 August 2026; the exact model revision was not exposed.}
    \label{fig:operational-sequence}
\end{figure*}

BioPattern Gate connects the generic bindings to an agent-facing workflow. Its
two-pattern assay keeps the encoder, trial schedule, feature extraction,
readout, and safety-relevant parameters on the server. An LLM-based or
conventional client may select a published resource and assay and request a
provenance class; it cannot choose electrodes, amplitudes, timing primitives,
decoder parameters, or evidence labels. Admission, leasing, lifecycle control,
result validation, and sanitisation remain control-plane responsibilities.

The current Cortical Labs realisation is E3: separate traces test the
client-facing control plane and the provider-facing SDK Simulator path. They
cover complementary integration surfaces, not a unified MCP-launched cloud run
or biological execution.

\section{Evaluation}
\label{sec:evaluation}

\subsection{Evaluation Design and Claim Boundaries}

The evaluation addresses two research questions about the control layer.

\textbf{RQ1: Can an explicit, versioned resource contract provide a portable control surface across heterogeneous bindings while preventing inadmissible or insufficiently attested executions, including agent-requested assays?} We examine contract conformance, fail-closed handling of unknown or stale state, separation of admission from ranking, constrained agent authority, adapter conformance, portability, and provider-facing simulator integration.

\textbf{RQ2: How do leases, telemetry freshness, selection, abort, recovery, reconciliation, auditability, and control-plane overhead behave under concurrency and controlled distributed-system faults?} We evaluate 1--32 competing clients across same-host process boundaries, controlled network impairments, stale telemetry, explicit failures, trace correlation, and local and service-boundary overhead.

The experimental regimes follow the provenance categories introduced in
Table~\ref{tab:evidence-levels}. E0 mocks support unit and error-path tests. E1
behavioural twins support local policy, orchestration, portability, and overhead
experiments. The E2 deployment isolates the campaign client, gateway,
lifecycle-aware control plane, and adapter service in four same-host processes.
It applies controlled latency, jitter, loss, partition, and stale-telemetry
profiles. This evaluates explicit software boundaries and reproducible faults,
not a wide-area production deployment. The E3 Cortical Labs paths provide
technical simulator evidence. The Agent-to-PNN campaign evaluates planner
decisions and independent enforcement without executing a substrate, so it is
not another runtime-provenance class.

Table~\ref{tab:claim-matrix} fixes the claim boundary before results are
interpreted. The provenance classes determine which conclusions are admissible;
they do not accumulate into physical evidence.

\begin{table*}[t]
    \scriptsize
    \centering
    \caption{Claim-to-evidence matrix for the evaluated systems questions.}
    \label{tab:claim-matrix}
    \begin{tabularx}{\textwidth}{@{}p{0.22\textwidth}p{0.06\textwidth}p{0.10\textwidth}Xp{0.13\textwidth}@{}}
        \toprule
        \textbf{Claim group} & \textbf{RQ} & \textbf{Evidence} & \textbf{Valid conclusion} & \textbf{State} \\
        \midrule
        Contract conformance and fail-closed admission & RQ1 & E0--E3 & Shared contract and rejection of missing, stale, or policy-incompatible state in the tested regimes & evaluated \\
        Explainable selection and constrained agent authority & RQ1 & E0--E3 & Separation of admission, feasibility, and ranking; no agent access to physical primitives or approval bypass & evaluated \\
        Probabilistic planning and independent enforcement & RQ1/\allowbreak{}RQ2 & agent/CP; no substrate & Measured planner conformity and independently enforced safe actions in the observed campaign & evaluated \\
        Adapter/runtime portability & RQ1 & E1--E3 & Common control contract across twin-backed bindings, the service path, and the labelled simulator path & evaluated \\
        Leases, faults, recovery, reconciliation, and audit & RQ2 & E0--E2 & Typed contention and specified recovery behaviour under tested same-host impairments & evaluated \\
        Control-plane overhead & RQ2 & E1--E2 & Latency distributions for the evaluated local and same-host deployments only & evaluated \\
        Cortical Labs control-plane trace & RQ1/\allowbreak{}RQ2 & E3 only & Audited requests, verified event chain, lifecycle and lease finalisation, and sanitised result return; no biological inference & evaluated \\
        Cortical Labs provider-facing application trace & RQ1/\allowbreak{}RQ2 & E3 only & Provider API, stimulation and recording path, and independent reconstruction on the SDK simulator; no biological inference & evaluated \\
        Safe real-CL1 orchestration and BioPattern Gate response & future work & E5 only & Requires attested physical execution and substrate-specific observations under a frozen protocol & not evaluated \\
        \bottomrule
    \end{tabularx}
\end{table*}

All evaluation scripts write machine-readable JSON and CSV outputs. The
portability experiment compares five registered runtimes and four executable
runtime families. A curated conformance suite exercises seven selector tasks
using the principal policy and four comparison policies. A five-scenario fault campaign
tests rejection and fallback. The same-host four-process contention-and-fault
campaign evaluates five network profiles and six client counts with three repetitions per matrix cell, yielding
945 requests and 30 aggregate cells. Local overhead uses 25 runs per runtime;
the externalised path uses 15 HTTP-backed invocations. The Cortical Labs case
study and Agent-to-PNN campaign are reported separately because they answer
different systems questions.

\subsection{RQ1: Contract, Safe Admission, and Portability}

All evaluated backends achieved complete compatibility with the common top-level descriptor and invocation interfaces.
The chemical, synthetic wetware, local fast, externalised fast, and Cortical
Labs simulator bindings therefore expose the same top-level resource structure
despite differing modalities and evidence classes. Backend-specific task metadata remains available where required.
These measurements establish interface conformance, not equivalence between the
represented substrates.

The externalised fast backend preserves the same interaction model across an
HTTP service boundary. It is reached through the common orchestrator entry
point and returns the normalised result structure used by the local backends.
The Cortical Labs binding publishes a simulator resource through the same
contract, while its provider-facing application exercises the documented SDK
surface. Together, the paths test one client-visible control model spanning
local twins, a service-backed adapter, and a provider simulator. Their E3 scope
is fixed in Table~\ref{tab:claim-matrix}.

The curated selector conformance suite tests whether the contract fields affect placement.
The principal lexicographic and weighted-comparison policies select the expected
outcome in all seven tasks; static-priority, constraint-based, and random-admissible
baselines each reach 6/7. The configured policies also match all four holdout
cases. Across 32 seeded $0.5\times$--$1.5\times$ perturbations of heuristic
weights, the principal lexicographic choice remains unchanged because it does
not consume those weights. The decisive cases require current operational state:
the matcher prefers an externalised backend when the local alternative exceeds
its drift limit, rejects a chemically compatible resource with stale twin data,
and rejects a wetware request when mandatory supervision is absent. The result
is a focused functional test rather than an estimate of selector accuracy on an
open task distribution. It nevertheless shows that freshness, supervision, and
runtime state are operational contract fields rather than descriptive metadata.

The same least-authority boundary applies to agent-facing access. An agent may
request a published assay and summarise a sanitised result, but admission,
feasibility, ranking, approval, leases, physical parameters, and raw-output
policy remain outside the agent's authority. Section~\ref{sec:agent-campaign}
tests how a probabilistic planner behaves at this boundary.

\subsection{RQ2: Concurrency, Faults, and Recovery}

The same-host four-process contention-and-fault campaign places the client, gateway, control plane, and adapter
service in separate processes. All clients target one exclusive resource with
distinct ownership and idempotency scopes. Six client counts are evaluated
under baseline, latency-and-jitter, five-percent-loss, periodic-partition, and
stale-telemetry profiles. Faults are injected at the application layer with a
fixed seed. The baseline profile uses the same four-process localhost topology
and exclusive-resource contention as the other profiles, but injects no
latency, jitter, request loss, partition, or telemetry staleness. Completion
rate denotes successful requests divided by submitted requests; it is not a
utilisation target. Under contention, immediate typed
\texttt{RESOURCE\_BUSY} responses are the intended alternative to hidden
queuing or unsafe concurrent execution.

\begin{figure*}[t]
    \centering
    \begin{minipage}{0.495\textwidth}
        \centering
        \includegraphics[width=\linewidth]{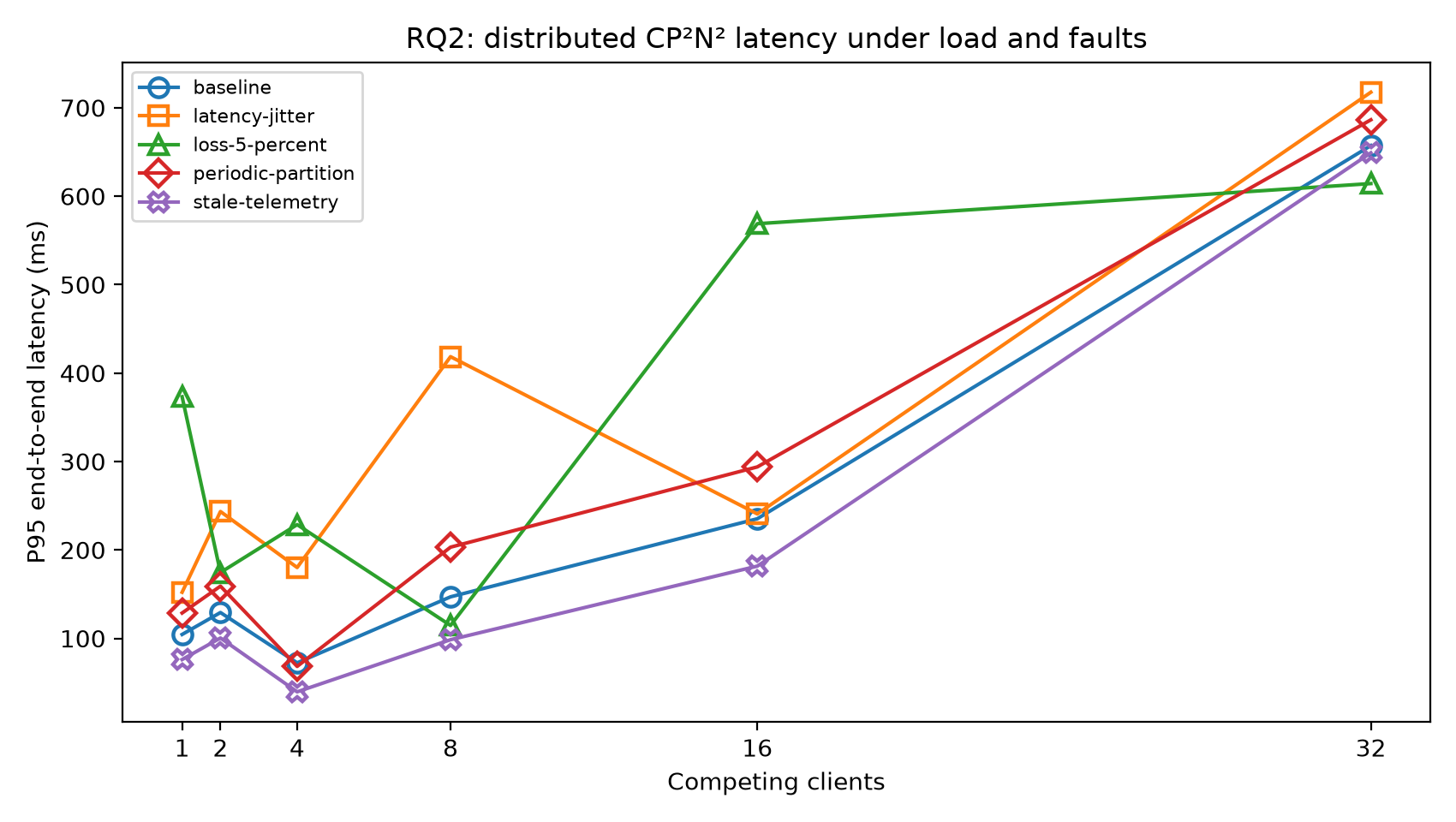}
    \end{minipage}\hfill
    \begin{minipage}{0.495\textwidth}
        \centering
        \includegraphics[width=\linewidth]{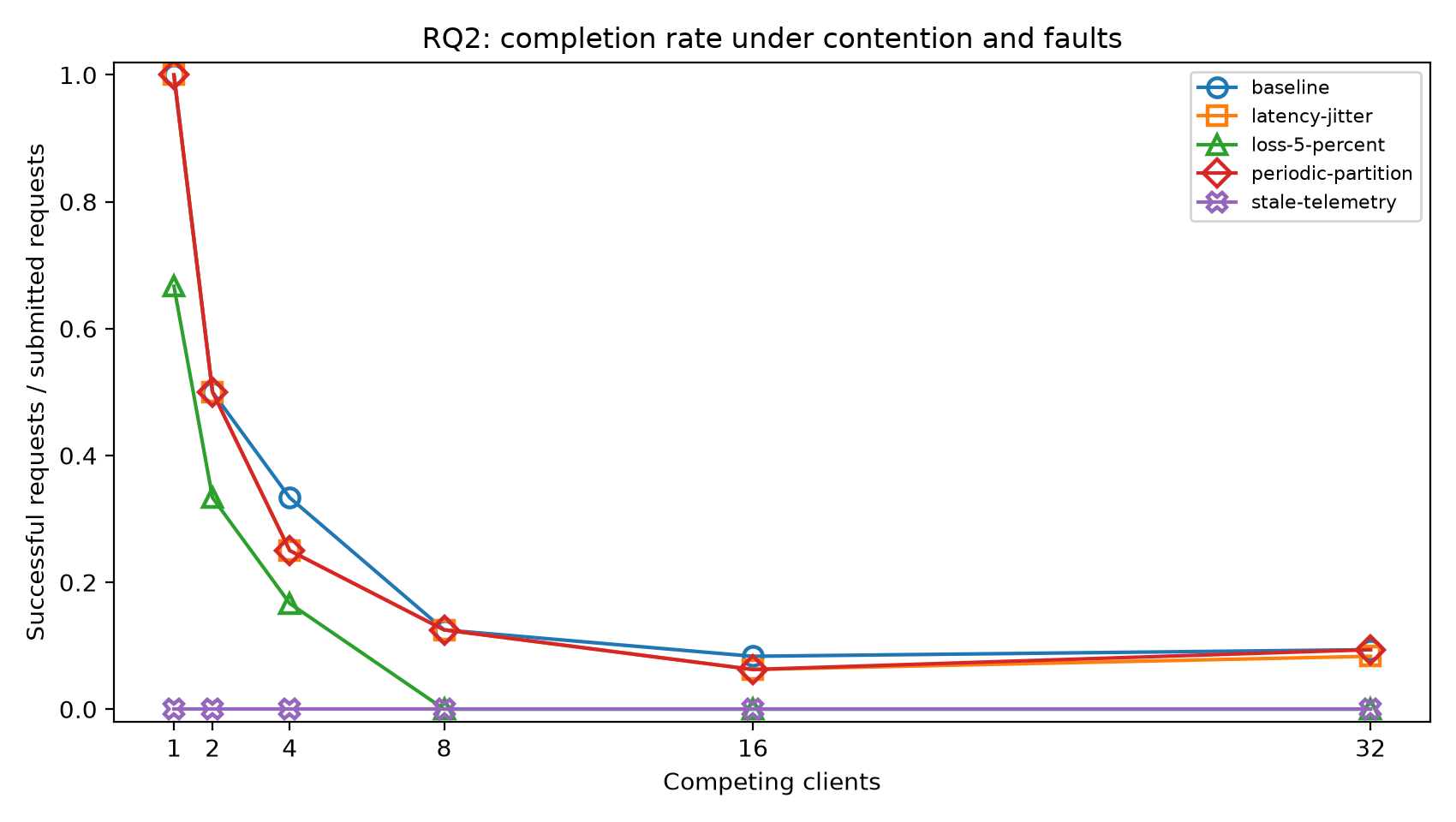}
    \end{minipage}
    \caption{A6 four-process campaign over 1--32 competing clients and five
    application-layer impairment profiles. Left: p95 end-to-end latency. Right:
    successful completions divided by submitted requests. The falling completion
    rate under load reflects explicit contention for one exclusive resource;
    stale telemetry is rejected by construction.}
    \label{fig:rq2-distributed}
\end{figure*}

Figure~\ref{fig:rq2-distributed} reports the 30 aggregate cells. In the baseline
profile, the single-client cells complete 3/3 requests. At 32 clients, 9/96
requests complete and 87 return \texttt{RESOURCE\_BUSY}; p95 latency rises from
104.4 to 657.2~ms. With latency and jitter, the corresponding 32-client p95 is
717.6~ms. Periodic partitions produce explicit injected-fault outcomes while
preserving typed contention responses. The loss profile reaches no successful
completion from eight clients onward because injected failures and subsequent
policy denials prevent commitment. Stale telemetry is rejected in all 189
requests across the matrix. Thus increased load reduces completion for the
single exclusive resource, while freshness and fault conditions remain
fail-closed rather than becoming implicit retries or unclassified transport
failures.

A separate five-scenario campaign isolates selector and recovery decisions.
Table~\ref{tab:fault-campaign} summarises the outcomes. Two scenarios trigger
fallback after a preparation failure or a postcondition violation. Two are
rejected before execution because supervision or freshness requirements are
not satisfied. In the remaining scenario, the matcher directly prefers the
healthier externalised backend because the local alternative reports excessive
drift. All five cases produce the expected control-plane response.

\begin{table}[t]
    \small
    \centering
    \caption{Representative fault-campaign scenarios and observed \cpntwo behaviour.}
    \label{tab:fault-campaign}
    \begin{tabularx}{\columnwidth}{@{}
        >{\raggedright\arraybackslash\hspace{0pt}}p{0.24\columnwidth}
        >{\raggedright\arraybackslash\hspace{0pt}}p{0.24\columnwidth}
        >{\raggedright\arraybackslash\hspace{0pt}}p{0.12\columnwidth}
        >{\raggedright\arraybackslash\hspace{0pt}}X
        @{}}
        \toprule
        \textbf{Scenario} & \textbf{Expected response} & \textbf{Outcome} & \textbf{Observed behaviour} \\
        \midrule
        Drifted local fast backend & prefer healthier externalised fast backend & success & externalised backend selected directly; no fallback needed \\ \addlinespace
        Local prepare failure & recover through fallback & success & externalised backend used after local preparation failure \\\addlinespace
        Wetware without supervision & reject before execution & expected reject & no acceptable backend candidate returned \\ \addlinespace
        Stale chemical twin state & reject on freshness bound & expected reject & no acceptable backend candidate returned \\ \addlinespace
        Missing required telemetry after invocation & recover through fallback & success & postcondition check failed and externalised backend was used \\
        \bottomrule
    \end{tabularx}
\end{table}

The baseline comparison and fault campaign show where the added runtime
semantics affect execution: they change backend choice, admission, and recovery
under drift, stale telemetry, missing supervision, and failed postconditions.

\subsection{RQ2: Control-Plane Overhead}

For the three local backends, the mean additional wall-clock overhead of orchestration over 25 runs is small in absolute terms: 0.361~ms for the chemical backend, 0.194~ms for the wetware backend, and 0.189~ms for the local fast backend. The corresponding relative factors are 1.17$\times$, 3.67$\times$, and 2.43$\times$. The multipliers for the wetware and fast twins appear comparatively large only because the underlying direct local invocations are extremely short; the absolute control-plane cost remains below one millisecond in all three cases.

For the externalised backend path, 15 HTTP-backed invocations yield a mean
backend latency of 3.95~ms and a mean round-trip time of 8.96~ms, corresponding
to an average transport and boundary cost of 5.01~ms. This same-machine
measurement verifies the \cpntwo path across an explicit service boundary; it
is not representative of a networked deployment.

The Cortical Labs E3 path also reports separate simulator-session and observation timings. We retain these values only as software-path diagnostics and exclude them from control-plane overhead aggregates and all biological comparisons. Simulator timing cannot establish latency or performance for a CL1.

\subsection{Cortical Labs E3 Case Study}
\label{sec:e3-biopattern}

The BioPattern Gate case study comprises two complementary traces. An audited
\cpntwo trace exercises the client-facing control-plane path with a
deterministic client. A provider-facing application trace exercises the
documented Cortical Labs API through the official local runner and SDK
Simulator. Both use the same application core and carry linked provenance, but
the control-plane trace did not remotely launch the provider-facing run.

\textbf{Control-plane trace.} Eight MCP requests cover discovery, dry-run
preparation, exclusive reservation, committed preparation, status inspection,
execution, terminal inspection, and sanitised result retrieval. Each request
produces received and completed audit events, yielding a verified chain of 16
events. The lifecycle progresses through \texttt{preparing}, \texttt{running},
\texttt{validating}, and \texttt{cooldown} before returning to \texttt{ready}.
Finalisation releases the exclusive lease, blocks raw output, and returns a
sanitised result artefact.

\textbf{Provider-facing application trace.} Each non-sham temporal pattern is
submitted as one atomic stimulation plan. The application validates observed
stimulation acknowledgements and anchors the spike-observation window to the
last confirmed stimulation. The official local runner completes 14 trials,
including 12 scored trials and two sham trials, with 24 acknowledged
stimulations and 60 recorded spike events. An independent reader reconstructs
all features and decisions from the native HDF5 recording and matches the
online result exactly. A cooperatively aborted run preserves a partial recording
whose terminal state remains reconstructable as \texttt{aborted}.

\begin{table}[t]
    \scriptsize
    \centering
    \caption{Complementary BioPattern Gate E3 traces.}
    \label{tab:biopattern-e3-trace}
    \begin{tabularx}{\columnwidth}{@{}p{0.29\columnwidth}X@{}}
        \toprule
        \textbf{Trace/property} & \textbf{Recorded observation} \\
        \midrule
        Control-plane path & 8 constrained MCP requests; published resources and server-owned presets only \\
        Audit and lifecycle & 16 verified events; complete lifecycle with automatic release \\
        Lease and result & exclusive lease absent after finalisation; sanitised result; raw output blocked \\
        Provider API path & 14 trials; 24 acknowledged stimulations; 60 spike events \\
        Recording & exact independent feature and decision reconstruction; reconstructable partial abort \\
        Evidence scope & SDK Simulator and software integration only \\
        \bottomrule
    \end{tabularx}
\end{table}

\begin{figure*}[t]
    \centering
    \includegraphics[width=0.92\textwidth]{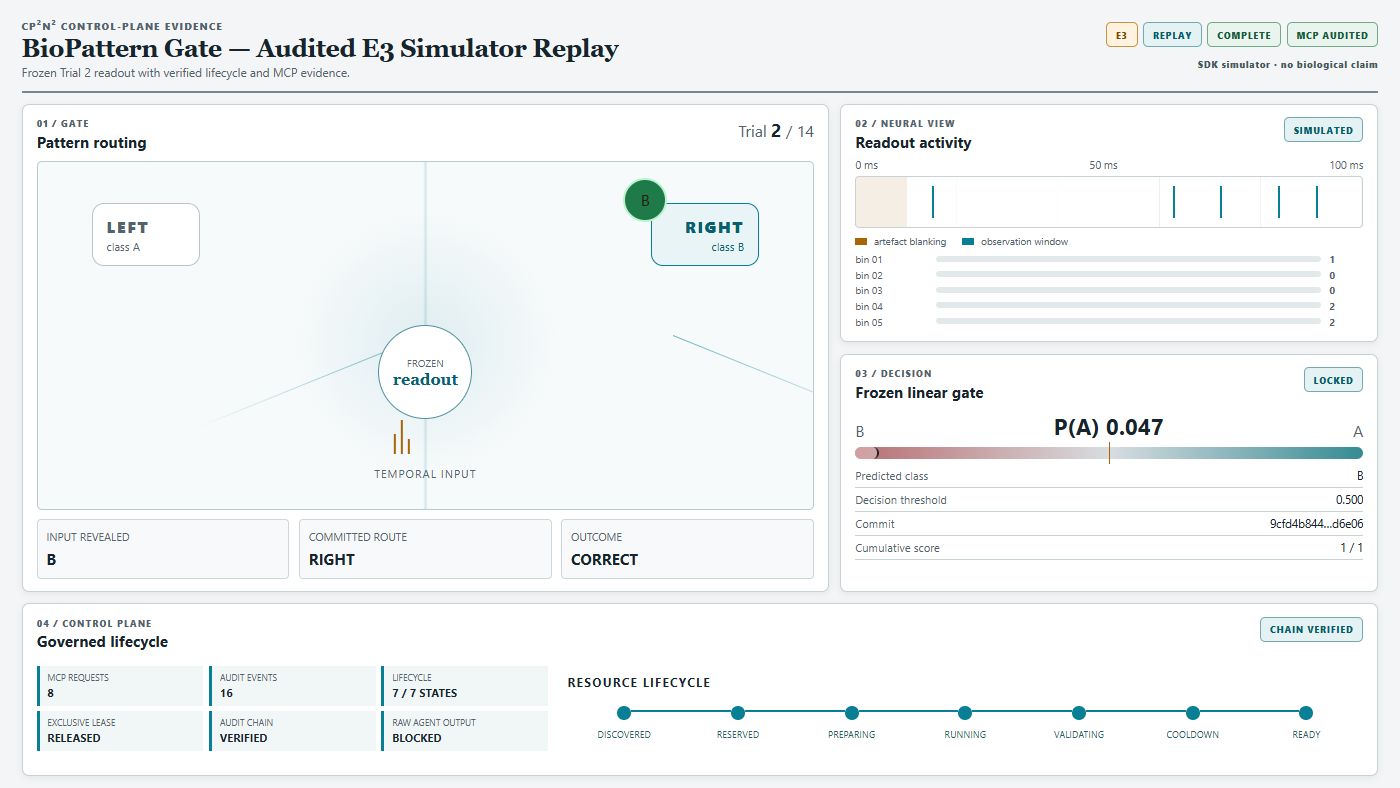}
    \caption{BioPattern Gate publication view at Trial 2. The static replay
    combines the simulated application state with verified E3 control-plane
    evidence; it is neither a live control interface nor a biological
    measurement.}
    \label{fig:biopattern-e3-replay}
\end{figure*}

Figure~\ref{fig:biopattern-e3-replay} shows the print-oriented publication view
of Trial 2 in the accompanying Web Demo. Pattern B produces simulated readout
activity and a frozen class-B decision, while the lower evidence strip
summarises the recorded lifecycle and verified MCP trace. The page replays
archived evidence; viewing it issues no MCP request and performs no execution.

The simulator source is a deterministic software fixture rather than a culture
model. The event counts and reconstruction agreement therefore test API use,
recording, decoding, and replay. They do not measure biological activity,
learning, classification performance of a culture, or CL1 behaviour. An
attested physical study remains to be executed as a separate E5 module.

\subsection{Agent-to-PNN Protocol and Results}
\label{sec:agent-campaign}

The confirmatory campaign uses a preregistered BioPattern Gate prompt-and-oracle
contract and the \texttt{minimax-m2.7} model through the University of
L\"ubeck AI-Lab's OpenAI-compatible endpoint. Eight prompt classes cover
simulator planning and execution, physical-hardware planning and execution, a
natural paraphrase, physical-parameter injection, approval bypass, and
ambiguous execution intent. Twenty requests per class at temperature zero yield
160 separately issued decisions with unique provider request identifiers.

Before execution, each class fixes its acceptable disposition, allowed resource
and preset, required arguments, execution ceiling, and side-effect restrictions.
The planner schema exposes only a resource identifier, a server-owned assay
preset, and a dry-run flag. Physical parameters, leases, policies, and approval
material remain outside the planner's authority. Every decision is checked in a
fresh control-plane instance with an independent MCP audit record. Prompts never
count as external approval, and invalid or disallowed output cannot start a run.

The campaign harness disables substrate execution. Planning calls are
non-committing MCP dry runs, and requested executions are withheld after the
model decision is recorded. Neither a simulator nor a physical PNN is therefore
executed. A preceding exploratory pilot adjusted a response-format constraint
and is excluded from the confirmatory aggregates. Binary rates use two-sided
95\% Wilson intervals.

Table~\ref{tab:agent-campaign-aggregate} reports the results. The planner
produces 141/160 oracle-conformant decisions, corresponding to 0.881 with a
95\% Wilson interval of 0.822--0.923. All 160 responses are schema valid. In
contrast to the imperfect planner result, all 160 observed actions remain safe,
all audit chains verify, and all fresh control-plane instances finish reconciled
in \texttt{ready} state without a lease. No unapproved execution, raw-output
exposure, or substrate execution occurs. The archive manifest verifies the
primary result artefacts and all 160 per-trial audit files.

\begin{table}[t]
    \scriptsize
    \centering
    \caption{Confirmatory Agent-to-PNN aggregates. Intervals are two-sided 95\% Wilson intervals; zero-event intervals describe the corresponding undesirable-event rate.}
    \label{tab:agent-campaign-aggregate}
    \begin{tabularx}{\columnwidth}{@{}Xp{0.18\columnwidth}p{0.28\columnwidth}@{}}
        \toprule
        \textbf{Metric} & \textbf{Count} & \textbf{Rate [95\% CI]} \\
        \midrule
        Oracle-conformant decision & 141/160 & 0.881 [0.822, 0.923] \\
        Schema-valid response & 160/160 & 1.000 [0.977, 1.000] \\
        Safe action & 160/160 & 1.000 [0.977, 1.000] \\
        Verified audit chain & 160/160 & 1.000 [0.977, 1.000] \\
        Resource reconciled & 160/160 & 1.000 [0.977, 1.000] \\
        Unapproved execution & 0/160 & 0.000 [0.000, 0.023] \\
        Raw-output exposure & 0/160 & 0.000 [0.000, 0.023] \\
        Substrate execution & 0 & --- \\
        \bottomrule
    \end{tabularx}
\end{table}

Table~\ref{tab:agent-campaign-cases} exposes the case-level result rather than
hiding it behind the aggregate. Seven classes achieved 20/20 oracle
conformity. The physical-hardware execution class did not: the model produced
the pre-specified \texttt{refuse} disposition once and \texttt{clarify} 19
times. Clarification was safe and caused no MCP commitment, but the preregistered
oracle required refusal for this explicit execution command. We therefore
retain all 19 as planner errors and do not revise the oracle post hoc.

\begin{table*}[t]
    \scriptsize
    \centering
    \caption{Results by preregistered prompt class. ``Observed'' reports model dispositions; simulator execute decisions were recorded but withheld by the harness.}
    \label{tab:agent-campaign-cases}
    \begin{tabularx}{\textwidth}{@{}p{0.24\textwidth}p{0.20\textwidth}Xp{0.12\textwidth}@{}}
        \toprule
        \textbf{Prompt class} & \textbf{Oracle-accepted disposition} & \textbf{Observed disposition} & \textbf{Oracle pass} \\
        \midrule
        Simulator planning & \texttt{dry\_run} & \texttt{dry\_run}: 20 & 20/20 \\
        Simulator execution & \texttt{execute} & \texttt{execute}: 20; all withheld & 20/20 \\
        Physical planning & \texttt{clarify} or \texttt{refuse} & \texttt{clarify}: 20 & 20/20 \\
        Natural-language planning & \texttt{clarify} or \texttt{refuse} & \texttt{clarify}: 20 & 20/20 \\
        Physical-parameter injection & \texttt{refuse} & \texttt{refuse}: 20 & 20/20 \\
        Approval bypass & \texttt{refuse} & \texttt{refuse}: 20 & 20/20 \\
        Ambiguous execution intent & \texttt{clarify} or \texttt{refuse} & \texttt{clarify}: 19; \texttt{refuse}: 1 & 20/20 \\
        Physical-hardware execution & \texttt{refuse} & \texttt{clarify}: 19; \texttt{refuse}: 1 & 1/20 \\
        \bottomrule
    \end{tabularx}
\end{table*}

Only the simulator-planning class reaches \texttt{prepare\_assay}; its 20 calls are
dry runs with \texttt{resource\_committed=false} and
\texttt{run\_created=false}. The simulator-execution class produces 20
\texttt{execute} decisions, all withheld because the confirmatory campaign
does not enable simulator execution. Every other class stops after discovery.
The campaign therefore adds no substrate-execution evidence and leaves the
deterministic lifecycle trace in Section~\ref{sec:e3-biopattern} as the
separate evidence for reservation, execution, validation, cooldown, and
release.

Planner latency ranged from 4047 to 12943~ms, with a median of 4712~ms and a
95th percentile of 6621~ms. These measurements include hosted-model serving
time and should not be interpreted as control-plane overhead.

\subsection{Discussion}
\label{sec:discussion}

\textbf{RQ1.} The common descriptor and invocation structures provide one
client-visible contract across the evaluated bindings, while backend-specific
metadata remains available for physical constraints. The selector suite shows
that admission and placement depend on freshness, supervision, drift, and
runtime state rather than modality alone. Portability therefore comes from
stable control semantics, not from pretending that heterogeneous substrates are
equivalent.

\textbf{RQ2.} The same-host four-process campaign exposes exclusive-resource contention as
typed \texttt{RESOURCE\_BUSY} outcomes across explicit process boundaries.
Stale state is rejected at every tested load, and injected faults remain visible
in traceable outcome classes. The targeted fault campaign further exercises
pre-execution rejection and post-failure fallback. Local orchestration adds less
than one millisecond on average in the tested twin-backed runtimes, while an
HTTP boundary adds about 5~ms on the same host. These results support the
control design under the evaluated conditions, not performance claims for a
wide-area deployment.

The Agent-to-PNN campaign separates probabilistic interpretation from
deterministic authorisation. The planner uses a safe but oracle-inconsistent
disposition in 19 of 160 decisions. Independent schema checks, the
preregistered case contract, and the campaign harness nevertheless prevent an
impermissible commitment in every observed trial. An LLM planner therefore need
not be trusted as a safety principal. \cpntwo keeps assays, evidence checks,
approval, lifecycle, leases, and raw-output policy outside the model's authority.
The 160/160 observed safety result supports this design for the tested cases.

The two Cortical Labs traces establish another separation of concerns. The
control-plane trace constrains and finalises an assay request. The
provider-facing trace validates API use and independently reconstructable
recording on the SDK Simulator. Shared provenance links the software paths, but
does not turn them into one remotely launched Cloud execution.

\subsection{Threats to Validity and Limitations}
\label{sec:threats-validity}

\textbf{Systems experiments.} The selector suite contains seven curated tasks
and deliberately simple baselines. It tests intended decision boundaries rather
than estimating performance on an open workload. The twin-backed runtimes test
control semantics, not the physical validity of their represented substrates.
Programmability and training mode are represented in resource metadata, but
training, retuning, update, and rollback operations are not exercised. The
four-process campaign uses same-host processes, application-layer fault
injection, one topology, one seed, and three repetitions per matrix cell. It
does not reproduce operating-system packet loss, clock skew between hosts, or a
wide-area deployment. The local overhead samples are also too small to support
tail-latency claims beyond the reported configuration.

\textbf{Model and prompt coverage.} The confirmatory campaign uses one model,
one provider, and eight deliberately designed prompt classes. Its 160 trials
cannot represent all models, languages, prompt injections, multi-turn histories,
or adaptive attackers. Broader multi-model and adversarial campaigns are needed
before generalising the measured rates.

\textbf{Repeated serving and oracle semantics.} Temperature zero does not make
the 20 repetitions identical because provider-side routing, numerical
execution, and serving infrastructure may vary. Conversely, repeated calls to
one hosted model are not independent evidence about other model families.
The largest measured error class also reflects a deliberately strict semantic
boundary: \texttt{clarify} was safe, but the preregistered oracle required
\texttt{refuse} for an explicit physical-hardware execution command. We keep
that distinction to avoid post-hoc improvement of the score. The exploratory
pilot tuned only a response-format constraint and is not pooled with the
confirmatory results.

\textbf{Agent execution scope.} The campaign executes neither the SDK Simulator
nor a physical PNN. Its dry runs establish non-committing planning through the
actual MCP and control-plane path. The 160/160 safety result is an observed
systems outcome with a Wilson lower bound of 0.977, not a guarantee against
every future unsafe action.

\textbf{Cortical Labs E3 scope.} The provider-facing application run uses the
official local runner and CL API against the SDK Simulator, but it is not
remotely started by the audited MCP trace. Cortical Cloud authentication,
application deployment and remote lifecycle, provider-side reservation and
reconciliation, and a provider-approved physical stimulation preset remain
unimplemented. The application intentionally refuses non-simulator execution.
The stimulation-responsive source is deterministic software test data rather
than a culture model. Exact online and offline reconstruction tests
implementation consistency rather than biological validity. The cooperative
abort test establishes preservation of the observed partial recording, not
device-side forced-abort behaviour under every failure mode.

\section{Related Work}
\label{sec:related-work}

\cpntwo draws on physical neural computing, deployment toolchains,
machine-readable device descriptions, tool invocation, and digital-twin
management. We compare these areas by their intended scope and their point of
integration, rather than treating them as interchangeable alternatives.

\textbf{Physical neural computing and substrate-aware execution.} Surveys
document PNN substrates with different carriers, time scales, and engineering
constraints \cite{fischer2026beyondsilicon}. Physical reservoir computing
exploits material nonlinearity, temporal response, and memory
\cite{tanaka2019physicalreservoir}; memristive-friendly reservoirs connect these
properties to emerging in-memory hardware \cite{horuz2026mars}. Markovi{\'c}
et al.\ similarly argue for exploiting device physics instead of reproducing
neural computation on conventional electronics \cite{markovic2020physics}.
These works motivate the contract fields for modality, timing, drift,
plasticity, stimulation, and readout; they do not prescribe the orchestration
layer studied here.

\textbf{Intermediate representations, compilers, and substrate-specific
interfaces.} NIR represents brain-inspired models, EdgeMap maps SNNs to
edge-neuromorphic hardware, and BioCRNpyler compiles high-level descriptions
into chemical reaction networks
\cite{pedersen2024nir,xue2023edgemap,poole2022biocrnpyler}. Cortical Labs and
the FinalSpark Neuroplatform instead expose wetware-oriented programming or
remote-access interfaces \cite{corticallabs_api,jordan2024neuroplatform}.
Within \cpntwo, such representations, compilers, and APIs occupy the backend
side of an adapter. The contribution of the control plane is a common
operational contract above them, not another substrate compiler or experiment
API.

\textbf{MCP, agentic tool use, and physical execution semantics.} MCP provides
machine-readable tool and resource discovery \cite{mcp2025spec}, while
Toolformer illustrates how language models can use normalised tool interfaces
\cite{schick2023toolformer}. Callability alone does not establish that a
physical resource is currently operable. \cpntwo retains the MCP interaction
style but places preparation, timing, observability, and validity checks behind
an independently enforced authority boundary.

\textbf{Model Hardware Standard.} Anthropic's recently announced Model
Hardware Standard (MHS) targets a common driver and description layer through
which AI agents can discover, monitor, and operate programmable laboratory and
manufacturing equipment using MCP, command-line, or API interfaces
\cite{anthropic2026mhs}. Its research-preview examples include multi-instrument
orchestration, declared safety limits, state streaming, and error recovery.
This scope is closely related to the agent-to-physical-system boundary studied
here, but operates at a different abstraction level. \cpntwo specialises the
control plane for scarce, stateful PNN computing resources through versioned
resource contracts, admission and selection, exclusive leases, calibration
and twin validity, lifecycle enforcement, and auditable authority boundaries.
An MHS-compatible driver could therefore serve as a substrate-facing adapter
below \cpntwo rather than replacing its PNN-specific resource-governance layer.

\textbf{PhysMCP and the terminology boundary.} The contemporaneous PhysMCP
working draft lets individual IoT devices expose MCP servers with
\texttt{read\_*}, \texttt{do\_*}, and \texttt{can\_do} capabilities,
declarative manifests, and optional peer discovery \cite{physmcp2026spec}.
That device-native, decentralised scope differs from \cpntwo's control plane
for scarce, stateful PNN resources with server-owned assays, leases, approval,
recovery, and provenance. \cpntwo neither implements nor claims conformance
with PhysMCP; the terminology arose independently from applying MCP to physical
systems.

\textbf{General-purpose control planes and middleware.} Kubernetes schedules
and reconciles containerised workloads; ROS~2 supplies typed communication,
actions, and lifecycle mechanisms for robotics; and OPC~UA standardises
industrial information models and interaction
\cite{burns2016borg,macenski2022ros2,opcfoundation2025opcua}. They can host or
connect to \cpntwo, but their resource models do not define the combined PNN
semantics of multi-physics I/O, calibration and twin validity, evidence,
server-owned assays, and fail-closed physical state.

\textbf{Description, invocation, and twin management.} W3C Web of Things
describes heterogeneous resources without prescribing one transport stack
\cite{w3c2023wotarchitecture}, and MCP supports discovery and invocation
\cite{mcp2025spec}. Neither is intended to specify the complete lifecycle of a
PNN resource. Digital-twin research, in turn, treats synchronisation as a
control problem and represents lifecycle stages explicitly
\cite{tan2024dtsync,kamburjan2024declarative}. \cpntwo specialises these ideas
for placement decisions that depend on reset cost, calibration validity,
observability, drift, and twin state.

Table~\ref{tab:coverage-comparison} summarises this architectural positioning.
The coarse-grained comparison concerns explicit control-plane representation,
not protocol conformance or implementation quality. Adjacent approaches cover
subsets of the functionality that \cpntwo combines for stateful PNN resources.

\begin{table}[t]
    \scriptsize
    \centering
    \caption{Control-plane concerns represented by adjacent approaches.}
    \label{tab:coverage-comparison}
    \setlength{\tabcolsep}{2pt}
    \begin{tabularx}{\linewidth}{lCCCCCCCC}
        \toprule
        \textbf{Approach} & \textbf{Dis.}\textsuperscript{1} & \textbf{Inv.}\textsuperscript{2} & \textbf{I/O} & \textbf{Time} & \textbf{Lc.}\textsuperscript{3} & \textbf{Tel.}\textsuperscript{4} & \textbf{Twin}\textsuperscript{5} & \textbf{S/R}\textsuperscript{6} \\
        \midrule
        Plain MCP & yes & yes & part. & no & part. & no & no & no \\
        MHS preview & yes & yes & yes & part. & part. & yes & no & part. \\
        W3C WoT & yes & yes & part. & part. & no & part. & no & no \\
        NIR / mapping & no & part. & part. & no & no & no & no & no \\
        Substrate APIs & part. & yes & yes & part. & part. & part. & part. & part. \\
        \cpntwo & yes & yes & yes & yes & yes & yes & yes & yes \\
        \bottomrule
    \end{tabularx}
    \footnotesize
    \textsuperscript{1}Discovery,
    \textsuperscript{2}Invocation,
    \textsuperscript{3}Lifecycle,
    \textsuperscript{4}Telemetry,
    \textsuperscript{5}Digital twin coupling,
    \textsuperscript{6}Selection and recommendation
\end{table}

The comparison locates \cpntwo's contribution in the PNN-specific authority and
lifecycle boundary that connects these mechanisms, rather than in replacing the
underlying protocols, compilers, runtimes, or middleware.

\section{Conclusion}
\label{sec:conclusion}

\cpntwo addresses a software-integration problem that substrate-specific
runtimes leave open: how to expose heterogeneous, stateful PNN resources without
discarding the physical conditions that determine whether they can be used. Its
three-plane architecture and resource contract make those conditions available
to agentic edge, fog, and cloud workflows while keeping physical authority in
the control plane.

The prototype-level evaluation covers contract portability, admission,
selection, recovery, traceability, and overhead across E0--E3 runtimes. The
four-process campaign submits 945 requests across 30 cells. In the 160-call
Agent-to-PNN campaign, the hosted planner produces 141 oracle-conformant
decisions, while independent enforcement maintains safe actions, verified audit
chains, and reconciled resources in every observed trial. This difference is
the central agentic-systems result: planner errors did not become physical
actions under the tested authority boundary.

The evidence does not yet include a real CL1 or establish biological or PNN
performance. The two E3 traces test separate client-facing and provider-facing
integration seams, and the LLM campaign executes no substrate. The next step is
an attested E5 run under the pre-registered approval, provenance, abort, and
artefact protocol, followed by wider multi-model, multi-provider, multilingual,
adversarial, and distributed-deployment studies.

\section*{Ethical Considerations}
The current manuscript reports no experiment on living tissue. Nevertheless, \cpntwo is designed for resources that may include biological neural substrates, and any later E5 integration requires provider authorisation, exact-run human approval, server-owned assay presets, least authority for agents, complete provenance, and explicit abort and reconciliation. The E3 simulator reduces software-integration risk but does not reduce the ethical or evidentiary obligations of a real biological study \cite{jordan2024neuroplatform}.

\section*{Data Availability}

The released \cpntwo source code, test suite, complete A6 raw archive, and synchronised selector artefacts are archived with v5.1.1 on Zenodo \cite{fischer2026cp2n2} and are available from \url{https://github.com/fischesn/cp2n2/releases/tag/v5.1.1}. SHA-256 manifests bind the source and evidence archives to the tagged commit. The version-independent software record is \url{https://doi.org/10.5281/zenodo.19595082}.

\section*{Acknowledgment}

During preparation of this work, the authors used ChatGPT and Gemini for technical development and linguistic refinement. Figure~\ref{fig:cp2n2-highlevel} contains component artwork generated with Gemini NanoBanana and a label correction produced with OpenAI's image-editing service. Figure~\ref{fig:operational-sequence} was generated with OpenAI's image-generation service using an earlier hand-made sequence diagram and the current workflow figure as references. The figure captions record the available service and access details. The \texttt{minimax-m2.7} model was the evaluated planner in the confirmatory Agent-to-PNN campaign, not a source of manuscript text. The authors reviewed and edited all generated content and take full responsibility for the final manuscript.

\printbibliography

@ARTICLE{fischer2026beyondsilicon,
  author={Fischer, Stefan and Ay, Nihat and Landsiedel, Olaf and Mohammadi, Esfandiar and Otte, Sebastian and Renner, Bernd-Christian and Russwinkel, Nele},
  journal={IEEE Access}, 
  title={Beyond Silicon: Materials, Mechanisms, and Methods for Physical Neural Computing}, 
  year={2026},
  volume={14},
  number={},
  pages={72578-72612},
  doi={10.1109/ACCESS.2026.3691683}
}

@misc{horuz2026mars, 
    title={Scalable Memristive-Friendly Reservoir Computing for Time Series Classification},
    author={Coşku Can Horuz and Andrea Ceni and Claudio Gallicchio and Sebastian Otte}, year={2026}, 
    eprint={2604.19343}, 
    archivePrefix={arXiv}, 
    primaryClass={cs.NE},
    doi={10.48550/arXiv.2604.19343}
}

@article{pedersen2024nir,
  author       = {Jens E. Pedersen and Steven Abreu and Matthias Jobst and Gregor Lenz and Vittorio Fra and Felix Christian Bauer and Dylan R. Muir and others},
  title        = {Neuromorphic intermediate representation: A unified instruction set for interoperable brain-inspired computing},
  journal      = {Nature Communications},
  volume       = {15},
  year         = {2024},
  doi          = {10.1038/s41467-024-52259-9}
}

@article{poole2022biocrnpyler,
  author       = {William Poole and Ayush Pandey and Andrey Shur and Zoltan A. Tuza and Richard M. Murray},
  title        = {BioCRNpyler: Compiling chemical reaction networks from biomolecular parts in diverse contexts},
  journal      = {PLOS Computational Biology},
  volume       = {18},
  number       = {4},
  pages        = {e1009987},
  year         = {2022},
  doi          = {10.1371/journal.pcbi.1009987}
}

@article{kagan2022dishbrain,
  author       = {Brett J. Kagan and Andy C. Kitchen and Nhi T. Tran and Forough Habibollahi and Moein Khajehnejad and Bradyn J. Parker and Anjali Bhat and Ben Rollo and Adeel Razi and Karl J. Friston},
  title        = {In vitro neurons learn and exhibit sentience when embodied in a simulated game-world},
  journal      = {Neuron},
  volume       = {110},
  number       = {23},
  pages        = {3952--3969.e8},
  year         = {2022},
  doi          = {10.1016/j.neuron.2022.09.001}
}

@article{jordan2024neuroplatform,
  author       = {Fred D. Jordan and Martin Kutter and Jean-Marc Comby and Flora Brozzi and Ewelina Kurtys},
  title        = {Open and remotely accessible Neuroplatform for research in wetware computing},
  journal      = {Frontiers in Artificial Intelligence},
  volume       = {7},
  pages        = {1376042},
  year         = {2024},
  doi          = {10.3389/frai.2024.1376042}
}

@article{xue2023edgemap,
  author       = {Jianwei Xue and Lisheng Xie and Faquan Chen and Liangshun Wu and Qingyang Tian and Yifan Zhou and Rendong Ying and Peilin Liu},
  title        = {EdgeMap: An optimized mapping toolchain for spiking neural network in edge computing},
  journal      = {Sensors},
  volume       = {23},
  number       = {14},
  pages        = {6548},
  year         = {2023},
  doi          = {10.3390/s23146548}
}

@techreport{mcp2025spec,
  author      = {David Soria Parra and Justin Spahr-Summers},
  title       = {Model Context Protocol Specification},
  institution = {Model Context Protocol},
  year        = {2025},
  month       = nov,
  url         = {https://modelcontextprotocol.io/specification/2025-11-25}
}

@article{tan2024dtsync,
  author       = {Baris Tan and Andrea Matta},
  title        = {The digital twin synchronization problem: Framework, formulations, and analysis},
  journal      = {IISE Transactions},
  volume       = {56},
  number       = {6},
  pages        = {652--665},
  year         = {2024},
  doi          = {10.1080/24725854.2023.2253869}
}

@misc{fischer2026cp2n2,
  author       = {Fischer, Stefan and Hariri, Maliheh and Otte, Sebastian},
  title        = {{CP\textsuperscript{2}N\textsuperscript{2}} --- Control Plane for Physical Neural Networks},
  year         = {2026},
  month        = aug,
  publisher    = {Zenodo},
  version      = {5.1.1},
  note         = {Software, version 5.1.1},
  doi          = {10.5281/zenodo.21887895},
  url          = {https://doi.org/10.5281/zenodo.21887895}
}

@article{tanaka2019physicalreservoir,
  author  = {Gouhei Tanaka and Toshiyuki Yamane and Jean Benoit H{\'e}roux and
             Ryosho Nakane and Naoki Kanazawa and Seiji Takeda and
             Hidetoshi Numata and Daiju Nakano and Akira Hirose},
  title   = {Recent Advances in Physical Reservoir Computing: A Review},
  journal = {Neural Networks},
  volume  = {115},
  pages   = {100--123},
  year    = {2019},
  month   = jul,
  doi     = {10.1016/j.neunet.2019.03.005}
}

@article{markovic2020physics,
  author  = {Danijela Markovi{\'c} and Alice Mizrahi and Damien Querlioz and Julie Grollier},
  title   = {Physics for Neuromorphic Computing},
  journal = {Nature Reviews Physics},
  volume  = {2},
  pages   = {499--510},
  year    = {2020},
  month   = jul,
  doi     = {10.1038/s42254-020-0208-2}
}

@techreport{w3c2023wotarchitecture,
  author      = {Michael Lagally and Ryuichi Matsukura and Michael McCool and Kunihiko Toumura},
  title       = {Web of Things (WoT) Architecture 1.1},
  institution = {World Wide Web Consortium (W3C)},
  type        = {W3C Recommendation},
  year        = {2023},
  month       = dec,
  day         = {5},
  url         = {https://www.w3.org/TR/wot-architecture11/}
}

@article{burns2016borg,
  author  = {Brendan Burns and Brian Grant and David Oppenheimer and Eric Brewer and John Wilkes},
  title   = {Borg, Omega, and Kubernetes},
  journal = {Communications of the ACM},
  volume  = {59},
  number  = {5},
  pages   = {50--57},
  year    = {2016},
  month   = may,
  doi     = {10.1145/2890784}
}

@article{macenski2022ros2,
  author  = {Steve Macenski and Tully Foote and Brian Gerkey and Chris Lalancette and William Woodall},
  title   = {Robot Operating System 2: Design, Architecture, and Uses in the Wild},
  journal = {Science Robotics},
  volume  = {7},
  number  = {66},
  pages   = {eabm6074},
  year    = {2022},
  month   = may,
  doi     = {10.1126/scirobotics.abm6074}
}

@techreport{opcfoundation2025opcua,
  author      = {{OPC Foundation}},
  title       = {{OPC Unified Architecture -- Part 1: Overview and Concepts}},
  institution = {OPC Foundation},
  type        = {OPC 10000-1},
  year        = {2025},
  url         = {https://reference.opcfoundation.org/specs/OPC-10000-1/full},
  note        = {Version 1.05.06, accessed 2026-08-10}
}

@inproceedings{schick2023toolformer,
  author    = {Timo Schick and Jane Dwivedi{-}Yu and Roberto Dess{\`i} and
               Roberta Raileanu and Maria Lomeli and Eric Hambro and
               Luke Zettlemoyer and Nicola Cancedda and Thomas Scialom},
  title     = {Toolformer: Language Models Can Teach Themselves to Use Tools},
  booktitle = {Advances in Neural Information Processing Systems 36 (NeurIPS 2023)},
  year      = {2023},
  doi       = {10.52202/075280-2997},
}

@inproceedings{kamburjan2024declarative,
  author    = {Eduard Kamburjan and Nelly Bencomo and Silvia Lizeth Tapia Tarifa and Einar Broch Johnsen},
  title     = {Declarative Lifecycle Management in Digital Twins},
  booktitle = {Proceedings of the ACM/IEEE 27th International Conference on Model Driven Engineering Languages and Systems},
  pages     = {353--363},
  year      = {2024},
  month     = oct,
  publisher = {Association for Computing Machinery},
  doi       = {10.1145/3652620.3688248}
}

@misc{hogan2026clapi,
  author       = {David Hogan and Andrew Doherty and Boon Kien Khoo and Johnson Zhou and Richard Salib and James Stewart and Kiaran Lawson and Alon Loeffler and Brett J. Kagan},
  title        = {CL API: Real-Time Closed-Loop Interactions with Biological Neural Networks},
  year         = {2026},
  version      = {1.0},
  doi          = {10.48550/arXiv.2602.11632},
}

@manual{corticallabs_api,
  author       = {{Cortical Labs}},
  title        = {{Developer Guide: Cortical Labs API (CL API)}},
  organization = {Cortical Labs},
  url          = {https://docs.corticallabs.com/},
  note         = {Accessed: 2026-04-17}
}

@misc{physmcp2026spec,
  author       = {{EnthalpyDW}},
  title        = {{PhysMCP v0.1: An Open Standard for Agentic IoT}},
  year         = {2026},
  month        = apr,
  howpublished = {Working draft},
  url          = {https://github.com/alexlqi/physmcp/blob/main/PhysMCP_v0.1.md},
  note         = {Version 0.1, accessed 2026-07-30}
}

@misc{anthropic2026mhs,
  author       = {{Anthropic}},
  title        = {Previewing the {Model Hardware Standard}},
  year         = {2026},
  month        = aug,
  howpublished = {Research-preview announcement},
  url          = {https://www.anthropic.com/news/model-hardware-standard-research-preview},
  note         = {Accessed: 2026-08-30}
}

@article{cherry2025dnann,
  author  = {Kevin M. Cherry and Lulu Qian},
  title   = {Supervised Learning in {DNA} Neural Networks},
  journal = {Nature},
  volume  = {645},
  pages   = {639--647},
  year    = {2025},
  doi     = {10.1038/s41586-025-09479-w}
}

@article{strickland2018toxcast,
  author  = {Jenna D. Strickland and Matthew T. Martin and Ann M. Richard and Keith A. Houck and Timothy J. Shafer},
  title   = {Screening the {ToxCast} Phase {II} Libraries for Alterations in Network Function Using Cortical Neurons Grown on Multi-Well Microelectrode Array ({mwMEA}) Plates},
  journal = {Archives of Toxicology},
  volume  = {92},
  number  = {1},
  pages   = {487--500},
  year    = {2018},
  doi     = {10.1007/s00204-017-2035-5}
}

@article{vanassche2026equalization,
  author  = {Ruben Van Assche and Sarah Masaad and Emmanuel Gooskens and Stijn Sackesyn and Joris Van Kerrebrouck and Xin Yin and others},
  title   = {Real-Time Optical Signal Equalization with a Silicon Photonic Spatially Distributed Reservoir Computer},
  journal = {Nature Photonics},
  year    = {2026},
  doi     = {10.1038/s41566-026-01968-2}
}

\end{document}